\documentclass[lettersize,journal]{IEEEtran}
\usepackage{cite}
\usepackage{booktabs}
\usepackage{enumerate}
\usepackage{graphicx}
\graphicspath{{figures}}
\usepackage[ruled, vlined, linesnumbered]{algorithm2e}
\usepackage{url}
\usepackage{comment}
\usepackage{amsmath}
\usepackage{amsthm}
\usepackage[outdir=./]{epstopdf}
\usepackage{subfigure}
\usepackage{float}
\usepackage{amsmath}
\usepackage{bm}

\newtheorem{Prp1}{\textbf{Property}}

\begin{document}

\title{Hyperflex: A SIMD-based DFA Model for Deep
Packet Inspection}

\author{Yang~Liu,
Wenjun~Zhu,~\IEEEmembership{Member,~IEEE},
         Harry~Chang,
         Yang~Hong,
         Geoff Langdale,
         Kun~Qiu,~\IEEEmembership{Senior Member,~IEEE},
         and~Jin~Zhao,~\IEEEmembership{Senior~Member,~IEEE} 
%\thanks{This work was supported by National Key R&D Program of China under grant No. 2022YFB3102902.}
 \thanks{Yang Liu, Wenjun Zhu, Kun Qiu, and Jin Zhao are with the School of Computer Science, Fudan University, Shanghai 200438, China }% <-this % stops a space
 \thanks{Harry Chang, Yang Hong and Geoff Langdale are with Intel Asia-Pacific Research \& Development Ltd., Shanghai 200240, China }}

%\markboth{IEEE TRANSACTIONS ON NETWORK AND SERVICE MANAGEMENT}%
%{Xu \MakeLowercase{\textit{et al.}}: Accelerating Deep Packet Inspection by SIMD-based Multi-literal Matching Engine}


\maketitle

\begin{abstract}
Deep Packet Inspection (DPI) has been extensively employed for network security. It examines traffic payloads by searching for regular expressions (regex) with the Deterministic Finite Automaton (DFA) model. However, as the network bandwidth and ruleset size are increasing rapidly, the conventional DFA model has emerged as a significant performance bottleneck of DPI. Leveraging the Single-Instruction-Multiple-Data (SIMD) instruction to perform state transitions can substantially boost the efficiency of the DFA model. In this paper, we propose Hyperflex, a novel SIMD-based DFA model designed for high-performance regex matching. Hyperflex incorporates a region detection algorithm to identify regions suitable for acceleration by SIMD instructions across the whole DFA graph. Also, we design a hybrid state transition algorithm that enables state transition in both SIMD-accelerated and normal regions, and ensures seamless state transition across the two types of regions. We have implemented Hyperflex on the commodity CPU and evaluated it with real network traffic and DPI regexes. Our evaluation results indicate that Hyperflex reaches a throughput of 8.89Gbit/s, representing an improvement of up to 2.27 times over Mcclellan, the default DFA model of the prominent multi-pattern regex matching engine Hyperscan. As a result, Hyperflex has been successfully deployed in Hyperscan, significantly enhancing its performance.
\end{abstract}

\begin{IEEEkeywords}
Deep Packet Inspection, Regular Expression, Deterministic Finite Automata.
\end{IEEEkeywords}

% \fi
\IEEEpeerreviewmaketitle

\section{Introduction}
\IEEEPARstart{D}{EEP} Packet Inspection (DPI)~\cite{bujlow2015independent,sherry2015blindbox,trabelsi2016network,ccelebi2023comprehensive} serves as a pivotal component in numerous network security frameworks, including Intrusion Detection/Prevention Systems (IDS/IPS)~\cite{SnortIDS,SuricataIDS,jamshed2012kargus}, Web Application Firewalls (WAF)~\cite{ModSecurity,Shorewall,muzaki2020improving}, and application identification systems~\cite{qiu2022traffic}. Unlike conventional packet inspection methods, which only focuses on the fixed 5 tuples in packet headers, DPI delves into the entire packet payloads\cite{7468531,sitaridi2016simd,BUJLOW201575,xu2024accelerating}. It analyzes the packet payloads by searching for specific rules (i.e., patterns or signatures), which are usually regular expression (regex) rules because of their enhanced expressiveness and ability to describe a wide variety of payload signatures~\cite{ficara2008improved,Sommer_Paxson_2003,thompson1968programming,yu2006fast}. Consequently, DPI necessitates extensive regex matching to fulfill its role. 

Nowadays, the network bandwidth is dramatically increasing~\cite{xu2023harry, 400GbpsEthernet, wg802ieee} and many DPI-based applications demand real-time stream processing, which takes the high demand of the regex matching efficiency~\cite{antonatos2004generating, bremler2014deep,ccelebi2023comprehensive}. However, the conventional regex engine struggles to meet the high-performance requirements for extensive and complex networks. It has been reported that RE2~\cite{GoogleRE2}, Google's regex engine, takes 4696 seconds to perform 1GB traffic matching with Intel Xeon Platinum 8180 CPU, while this traffic contains only 818,682 packets~\cite{wang2019hyperscan}. Another popular regex engine, PCRE~\cite{PCRE2}, is even slower than it. DFA (Deterministic Finite Automata) is frequently utilized in regex matching due to its stable processing and no-backtracking characteristic, ensuring efficient and consistent regex recognition~\cite{el2017survey,kumar2006advanced,qiu2021teddy,xu2024accelerating,10.1145/3232195.3232201}. So it is an urge to enhance the performance of the DFA model. There have been some research efforts towards improved DFA models~\cite{wang2019hyperscan,zhu2022hyperverse}.

The well-known regex matching engine, Hyperscan~\cite{wang2019hyperscan}, supports parallel matching of regex rules by employing graph decomposition techniques to transform regex matching into a series of string matching and automaton matching tasks. It strives to employ DFA model for subsequent state transition processes to fully exploit the high-performance advantages of DFA. Hyperscan boasts performance enhancements of up to 10.3x over RE2 and 2.3x improvement over PCRE2, positioning it as the fastest regex engine~\cite{wang2019hyperscan}.The DFA model of Hyperscan is called Mcclellan, which leverages efficient state analysis algorithms to simplify state graphs to facilitate efficient state transitions. However, Mcclellan still employs the traditional DFA state transition algorithm, which is fundamentally a memory access procedure on a two-dimensional table. This method inherently incurs significant time overhead due to the necessity of searching the state transition table, where each transition may be hindered by memory access latency. As such, there exists considerable potential for enhancing the performance of the traditional state transition algorithm.

The recently proposed high-performance DFA model Hyperverse\cite{zhu2022hyperverse} uses SIMD instructions to accelerate DFA state transitions and incorporates a parallel state transition algorithm. Compared to the traditional DFA model, Mcclellan, Hyperverse significantly improves performance. However, this model relies on SIMD instructions, and most contemporary CPUs are equipped with SIMD instruction sets that do not surpass 512 bits in length, such as the AVX512 instruction set, thereby limiting the maximum number of DFA states to 64. So Hyperverse is only applicable to DFAs with fewer than 64 states. Moreover, numerous real-world DPI regexes, when transformed into DFAs, involve a state count significantly exceeding 64 states~\cite{ficara2008improved}. Thus Hyperverse cannot be effectively applied to practical matching scenarios. 

In this paper, we introduce Hyperflex, a SIMD-based DFA model that enables all regexes to benefit from the high performance of SIMD instructions. Initially, Hyperflex identifies the most suitable region from the whole regex to apply with SIMD-based parrel state transition. The selection of the region significantly impacts the efficiency of the model. Next, we design a hybrid transition algorithm to deal with the state transition of two kinds of regions and ensure seamless state transitions across different regions. Notably, Hyperflex reaches a throughput of 8.89 Gbit/s, up to 2.27x of Mcclellan. It has been successfully deployed in Hyperscan. Briefly speaking, this paper makes the following contributions:
\begin{itemize}
    \item we propose a region detection algorithm, which detects the most suitable region of the entire state graph for employing SIMD-based state transition. When a state enters a suitable region, it is more likely to circulate within the region rather than exiting it. This can take advantage of the high-performance of SIMD instruction as soon as possible. The algorithm substantially enhances the performance of our model. If the region is chosen randomly, the performance of Hyperflex even decreases by 1.66x.
    \item We design a hybrid transition algorithm to make states from different regions run under different mechanisms, while meticulously managing the transitions across these regions. As part of this algorithm, we design the gutter state transition table to address the potential mismatching issues caused by region switching. Moreover, to expedite the identification of the earliest escape state, we incorporate a SIMD-based escape state detection method into the hybrid algorithm. These components operate in concert to ensure seamless state transitions and enhance efficiency.
    \item We implement Hyperflex on a commodity CPU with AVX512 SIMD support. We conduct experiments to compare Hyperflex with Mcclellan and other mainstream DFA models.
\end{itemize}
% % You must have at least 2 lines in the paragraph with the drop letter
% % (should never be an issue)
% I wish you the best of success.

 

%\IEEEpubidadjcol

\section{related work}
\subsection{Deterministic Finite Automaton}
%Before we give detailed information on the state transition algorithm, we introduce the primitive DFA.
A deterministic finite automaton (DFA) is defined as a 5-tuple$\begin{Bmatrix} S, \Sigma, M,s_0, F\end{Bmatrix}$, which embodies: a finite collection of states $S$, an alphabet $\Sigma$ constituting the set of input symbols, a transition function $M$, a designated initial state $s_0$, and a collection of accept states $F$~\cite{gribkoff2013applications,vayadande2022simulation,fernau2015multi}.  The matching procedure is as follows. Consider $c=\{c_0,c_1, c_2, \ldots, c_n\}$ as a string formulated from the alphabet $\Sigma$. Typically, a DFA initializes with the state $s_0$. Then the DFA progresses through each character of string $c$, it transitions from state $s_{i-1}$ to state $s_i$ by the transition function $M$. If the final state $s_n \in F$, then the string is accepted by the DFA. Conversely, the DFA rejects the string, culminating in a non-match~\cite{ron1995learning,holub2009parallel,kumar2006advanced}. We conceptualize the operation of state transition with $T$. The process of DFA can be expressed as Formula 1:\\
\begin{equation}
s=T[c][s]\end{equation}
Formula 1 shows that the performance of DFA is not only affected by the number of states but also determined by the performance of the state transition algorithm.
\subsection{Hardware Acceleration Methods}
To accelerate regex matching, there have been some solutions leveraging hardware accelerators like FPGA, GPU or TACM\cite{CPUFPGA,5959160,MSDFA,Impala,NovelHardware,ScalablePatternMatching,RegexMatchingFPGA,lin2009compact,vasiliadis2009regular,meiners2011split}.

For example, Lin et al.\cite{lin2009compact} devised an FPGA-based acceleration architecture, CPDFA, which employs a mixed storage strategy to reduce DFA memory requirements: high-frequency states are stored via indirect indexing, while low-frequency states utilize direct indexing. Meiners et al.\cite{meiners2011split} introduced a TCAM-based DFA optimization scheme using dimension splitting. This method decomposes high-dimensional classifiers into multiple lower-dimensional classifiers and distributes them across the TCAM pipeline, lowering memory consumption while boosting throughput.

Nonetheless, these methods exhibit three main drawbacks: high capital costs, insufficient memory to hold a large number of rules and limited flexibility, as any algorithmic update necessitates a complete hardware redesign.
Although performance does improve, these solutions are rarely deployed. It is still
the software regex matching algorithms running on CPUs that
are adopted in most real scenarios.
\subsection{DFA Space Compression Algorithms}

Kong et al.\cite{kong2008efficient} observe the similarity in the transition behaviors of different input characters in the DFA transition table and partition the state set into multiple disjoint subsets. For each subset, they create an independent re-encoded alphabet and map the corresponding transition table to this new encoding through a mapping table, thereby reducing the overhead of redundant storage.

Kumar et al.\cite{kumar2006advanced} propose the classic D\textsuperscript{2}FA transition edge compression algorithm. The core idea behind this algorithm is to share transitions among states with similar transition tables and to establish default transitions between them, thus reducing memory usage. However, default transition paths may grow excessively long, which in turn raises memory access overhead. To address this issue, Becchi et al.\cite{becchi2013dfa} introduce the A-DFA method. By imposing a “state depth” constraint, A-DFA restricts the number of default-transition hops, ensuring that each character visits at most two states. Under identical memory bandwidth constraints, A-DFA exhibits a significantly higher compression rate than D\textsuperscript{2}FA.

DFA space compression algorithms can reduce DFA space consumption but often incur substantial preprocessing time, making them unsuitable for dynamic network scenarios with large-scale rule sets and frequent rule updates. The practical DFA performance is greatly affected of the state transition algorithm.

\subsection{The Traditional DFA State Transition Algorithm}
\begin{figure}[htb]
\centerline{\includegraphics[width=0.5\textwidth]{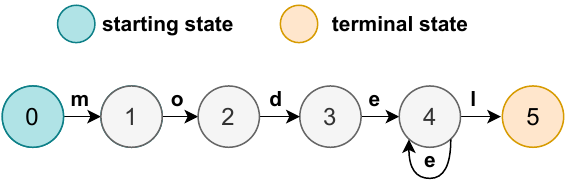}}
\caption{DFA representation for regex ``mode+l''}
\label{fig_dfarepre}
\end{figure}

The Traditional DFA State Transition Algorithm relies on a two-dimensional transition table for navigating state transitions. This table is meticulously structured, with rows corresponding to states generated from the regex, and columns indexed from 0 to 255, encompassing all the 256 ASCII characters.  
This structured approach ensures that the DFA model, from any specific state, can accurately ascertain the next state by querying the table according to the input character.  For instance, considering the regex ``model'', Fig.~\ref{fig_dfarepre} visualizes its state graph, while Table \ref{tab:1} details the transition table extracted from this graph, where ``5'' indicates the accept state.

\begin{table}[t!]
  \centering
  \caption{Traditional State Transition Table for Regex ``mode+l''}
  \setlength{\tabcolsep}{5pt} % 调整列间距
  \renewcommand{\arraystretch}{0.7} % 调整行间距
  \resizebox{0.45\textwidth}{!}{
    \begin{tabular}{c|c|c|c|c|c|c|c|c|c}
    \toprule
    \textbf{State} & ... & d & e & ... & l & m & ... & o & ... \\
    \midrule
    0 & ... & 0 & 0 & ... & 0 & 1 & ... & 0 & ... \\
    \midrule
    1 & ... & 0 & 0 & ... & 0 & 0 & ... & 2 & ... \\
    \midrule
    2 & ... & 3 & 0 & ... & 0 & 0 & ... & 0 & ... \\
    \midrule
    3 & ... & 0 & 4 & ... & 0 & 0 & ... & 0 & ... \\
    \midrule
    4 & ... & 0 & 4 & ... & 5 & 0 & ... & 0 & ... \\
    \midrule
    5 & ... & 0 & 0 & ... & 0 & 1 & ... & 0 & ... \\
    \bottomrule
    \end{tabular}
  }
  \label{tab:1}
\end{table}

\begin{figure}[htb]
\centerline{\includegraphics[width=0.5\textwidth]{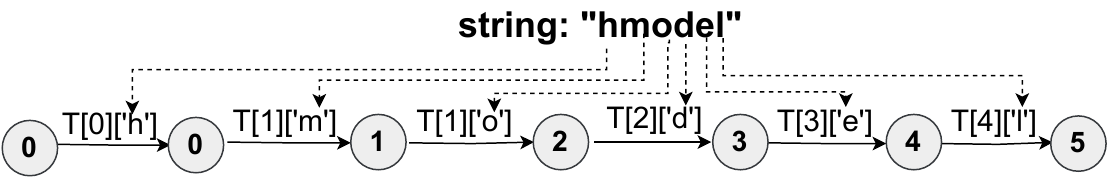}}
\caption{Traditional state transition algorithm working progress}
\label{fig_tt}
\end{figure}

The working process of the traditional state transition algorithm is essentially a sequence of queries to the state transition table. The transition to the subsequent state is determined by the current state and the character in the string. Consider the example string ``hmodel'', the matching progress is illustrated in Fig.~\ref{fig_tt}. This matching mechanism is straight, and the table's capacity to encompass all possible states enables the DFA model to apply to all regexes.  However, if the table consists of 500 states that fit into the L1 (the latency is one CPU cycle) cache. Because the  L1 cache size is typically under 64KB, it cannot accommodate the extensive size of the transition table, which may approximate 500 KB.  Consequently, each step in the matching process likely involves loading segments of the table from main memory into the L1 cache, incurring a latency of 4 to 5 CPU cycles. It also requires 2 cycles of arithmetic computation to get the position within the transition table. Therefore, the time spent on each matching process amounts to 7 to 8 CPU cycles, presenting substantial potential for optimization.

\subsection{The SIMD-based DFA State Transition Algorithm}
\begin{figure}[htb]
\centerline{\includegraphics[width=0.3\textwidth]{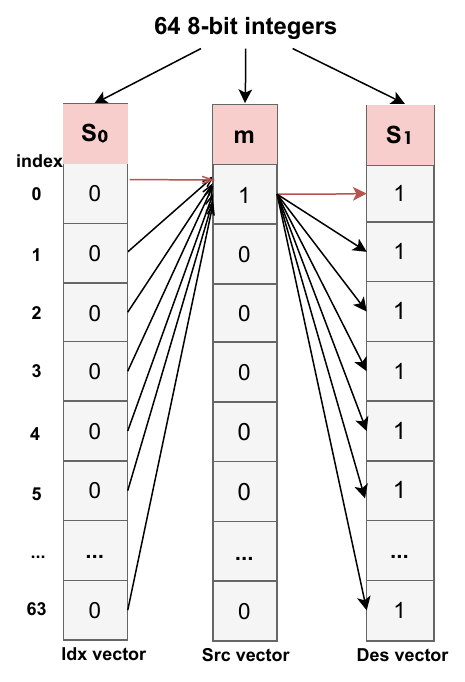}}
\caption{Shuffle operation for state transition}
\label{fig_vpermb}
\end{figure}

Hyperverse is a SIMD-based DFA model~\cite{zhu2022hyperverse}, which introduces a novel SIMD-based algorithm for state transitions. It enables an integer state into a vector comprising 64 identical state numbers and utilizes the SIMD instruction shuffle for the transition process, eliminating the need to directly access the transition table.  Fig.~\ref{fig_vpermb} shows how two state vectors perform a shuffle operation for state transition. The shuffle operation rearranges 8-bit integers in \textit{Src} across lanes using the corresponding indices in \textit{Idx}, and stores the result in \textit{Des}.  The ``$S_0$'' is a start vector of state, and the \textit{``m''} is the state vector of the ASCII `m'.  Through the shuffle operation, the model can take out the value in the input character `m' vector according to the value of the initial state to get the subsequent state ``$S_1$''.

\begin{table}[t!]
  \centering
  \caption{SIMD-based State Transition Table for Regex ``model''}
  \setlength{\tabcolsep}{5pt} % 调整列间距
  \renewcommand{\arraystretch}{0.6} % 调整行间距
    \begin{tabular}{c|c}
    \toprule
   \small\textbf{Character} & \small\textbf{Vector (label 0--63)} \\
    \midrule
    ... & ... \\
    \midrule
    d & \{003000...0\} \\
    \midrule
    e & \{000440...0\} \\
    \midrule
    ... & ... \\
    \midrule
    l & \{000050...0\} \\
    \midrule
    m & \{100000...0\} \\
    \midrule
    ... & ... \\
    \midrule
    o & \{020000...0\} \\
    \midrule
    ... & ... \\
    \bottomrule
    \end{tabular}
  \label{tab:2}
\end{table}

\begin{figure}[htb]
\centerline{\includegraphics[width=0.4\textwidth]{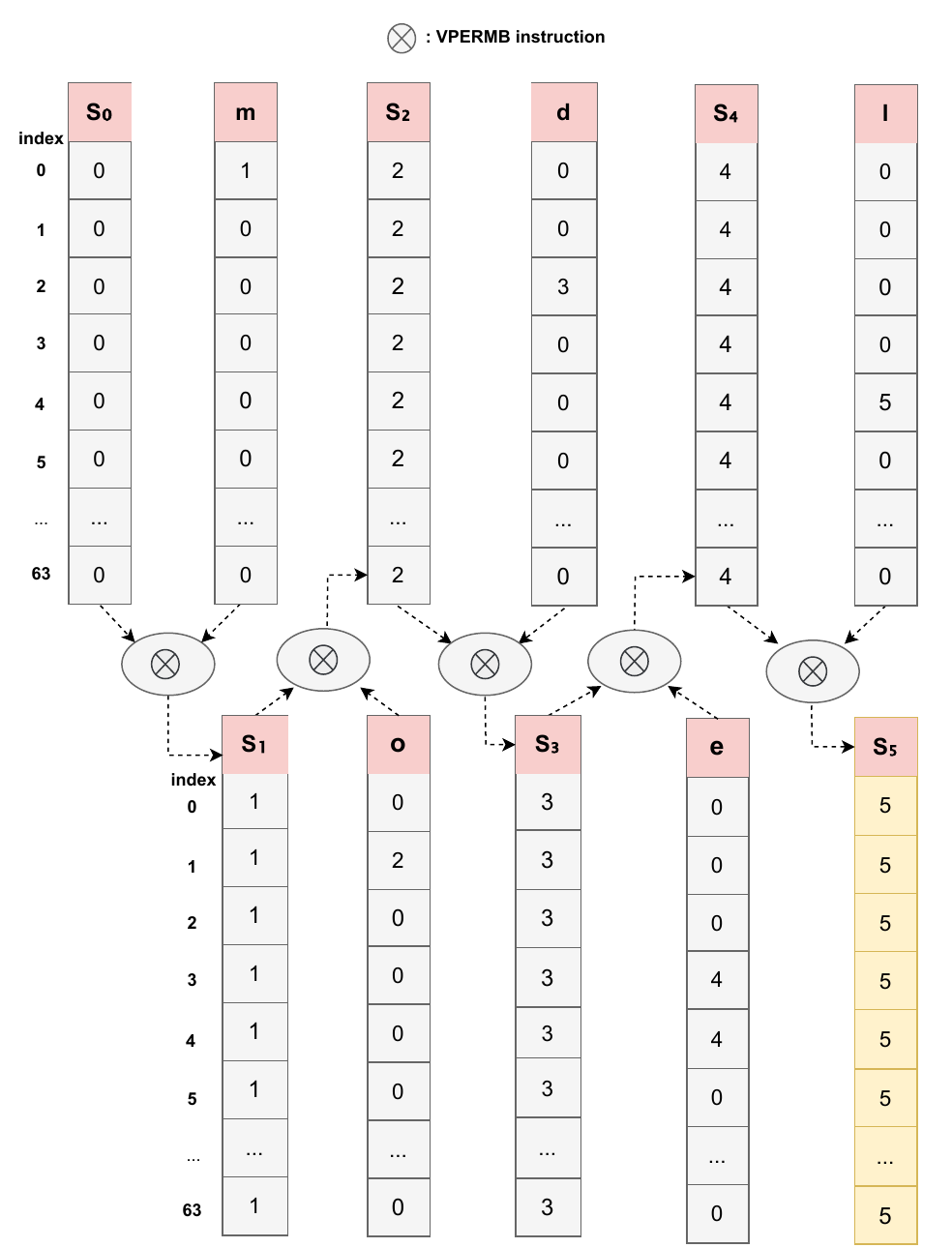}}
\caption{SIMD-based state transition algorithm working progress}
\label{fig_hyperset}
\end{figure}

We use the input string ``model'' and the regex ``mode+l'' as an example to explain the whole matching progress in Fig.~\ref{fig_hyperset}. For each character in the input string, we retrieve a corresponding vector from Table \ref{tab:2}, derived from the state graph. This table encompasses vector representations of all ASCII characters, with each vector being 512 bits and comprising 64 integer states. The size of the table is 16KB, which can be easily accommodated by the L1 cache. So it saves the time that loads the table from memory to the L1 cache. Then, we operate a shuffle operation with the character vector and the state vector to get the subsequent state vector, which consumes 3 CPU cycles. Finally, we also get the correct accept state. The process leverages the shuffle instruction as an alternative to traditional table queries, consuming merely 4 CPU cycles. 

Hyperverse further exploits the parallelism of SIMD instructions. After introducing SIMD instructions, two characters' corresponding SIMD vectors can take the VPERMB instruction to progress the state transition, which does not depend on the previous state. Thus, the original serial operation can be split into two parallel chains, thereby further accelerating the state transition speed.

We also take the input string ``model'' as an example and decompose its matching process into two chains, as shown in Equation~\ref{111}. ``$S_{init}$'' denotes the initial state vector, while ``$m$'' and other letters represent the SIMD vectors corresponding to the input characters. ``$S \otimes m$'' denotes performing the VPERMB operation on ``$S$'' and ``$m$''.

\begin{equation}
\label{111}
S_{model} = \{S_{init} \otimes m \otimes o\} \otimes \{d \otimes e \otimes l\}
\end{equation}

The result of Chain 2, $\{d \otimes e \otimes l\}$, is independent of the result of Chain 1, $\{S_{init} \otimes m \otimes o\}$, and can be executed in parallel. By dividing the entire state transition process into two computational chains, the serial nature of the state transitions is replaced with a parallel model, further improving the speed of state transitions. This design further enhances Hyperverse’s performance, making it the fastest DFA model available today. However, it only applies to regex containing less than 64 states, given that the AVX512 instruction spans 512 bits, accommodating a maximum of 64 states\cite{kusswurm2022simd,zhou2002implementing}. So Hyperverse can not be applied in practical scenarios as numerous regexes transformer to DFAs comprise states more than 64.

\section{HYPERFLEX ARCHITECTURE}
\subsection{Design Overview}\label{AA}
\begin{figure*}[htb]
\centerline{\includegraphics[width=1.0\textwidth]{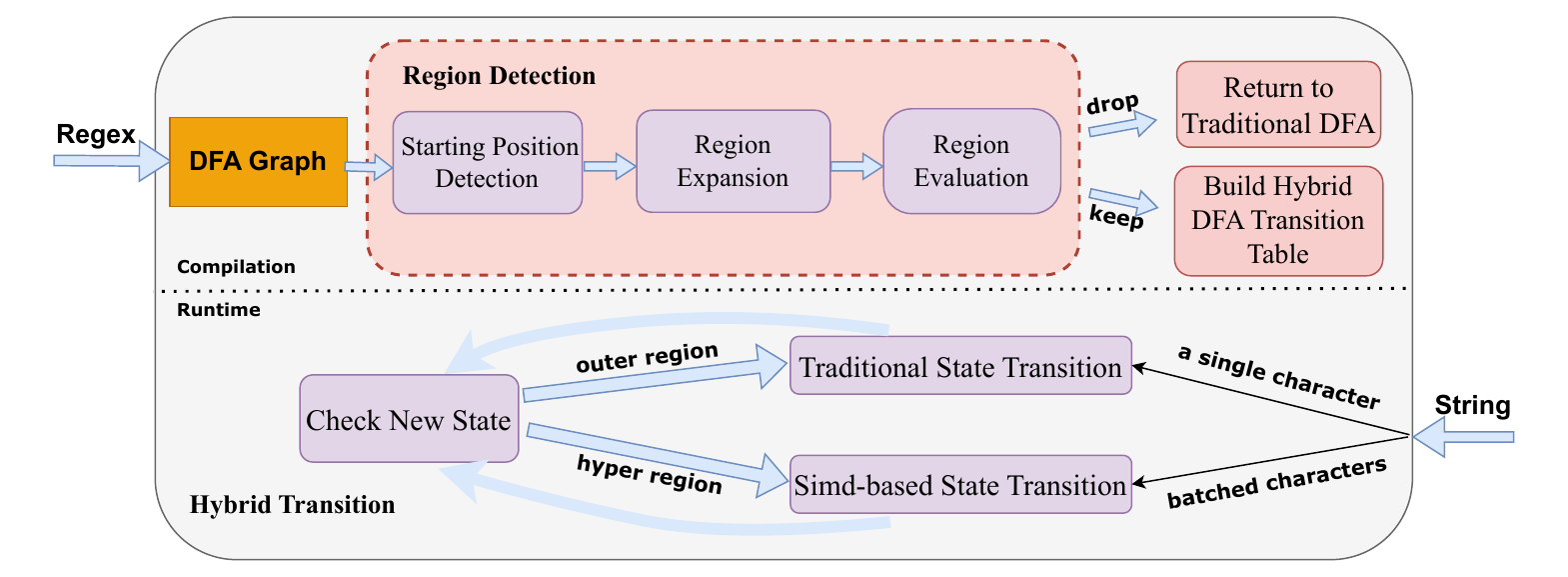}}
\caption{Design overview for the Hyperflex DFA model}
\label{fig_overview}
\end{figure*}

As shown in Fig.~\ref{fig_overview}, Hyperflex contains two main stages: the compilation stage and the runtime stage. In the compilation stage, the input regex is transformed into a DFA graph where nodes represent states and edges represent state transitions. Then, for applying the SIMD-based state transition algorithm, a suitable region limited to 64 states is identified from the whole graph. If the region is accepted via evaluation, the inter-region transition table and the outer-region transition table are constructed to facilitate the subsequent matching process. In the runtime stage, the current state is examined to determine which region it is in and decide to use which transition table to process the state transition. A state transition is executed based on the current state, the relevant transition table, and the subsequent character in the input string.
In the matching process, it is crucial to ensure the continuity of state transitions across regions. This system makes Hyperflex achieve better state transition performance than traditional DFA model and meanwhile solves the scale capability issue of the Hyperverse.

\subsection{The Region Detection Algorithm}
The SIMD-based state transition algorithm is more efficient but only applies to DFA whose states are fewer than 64, owing to the limitation of the longest SIMD vector in modern CPUs. To employ the algorithm with any regex, it is imperative to select a suitable state region from the entire DFA graph, referred to as the \textbf{hyper region}. The rest of the DFA graph is called \textbf{outer region}. Detecting the hyper region necessitates a comprehensive analysis of the entire graph to determine the starting state, followed by expansion to get the hyper region. Subsequently, the quality of the hyper region is evaluated to determine acceptance or rejection.  It is crucial to pick a suitable region in our design, which significantly impacts the performance of Hyperflex. The detail of the region detection algorithm is shown in Part IV.

\subsection{The Hybrid Transition Algorithm}
The hybrid transition algorithm makes states from different regions run under different mechanisms. It examines the current state's location to determine the next transition process. In the state transitions within the hyper region, to reduce the frequency of checking the region to which a state belongs and to leverage the parallelism of SIMD instructions, multiple characters are processed in a batch for parallel state transitions. Additionally, to address the issue of a state exiting the region within the same batch, which could lead to subsequent matching errors, we propose a novel state transition table. This new table works in conjunction with the original state transition table to accurately detect instances where a state exits the region and ensures proper handling of subsequent processes. Detailed information about the hybrid transition algorithm is provided in Part V.

% There are many grammar mistakes I am going to revise
\section{The Hyper Region Design}
The core of the Hyperflex design is to find the hyper region of the whole DFA graph to apply the SIMD-based state transition algorithm. This is a new problem both in automata theory and graph theory. We define the problem on our own and provide an analytical modeling framework.
\subsection{Problem Analysis}\label{AA}
\begin{figure}[htb]
\centerline{\includegraphics[width=0.5\textwidth]{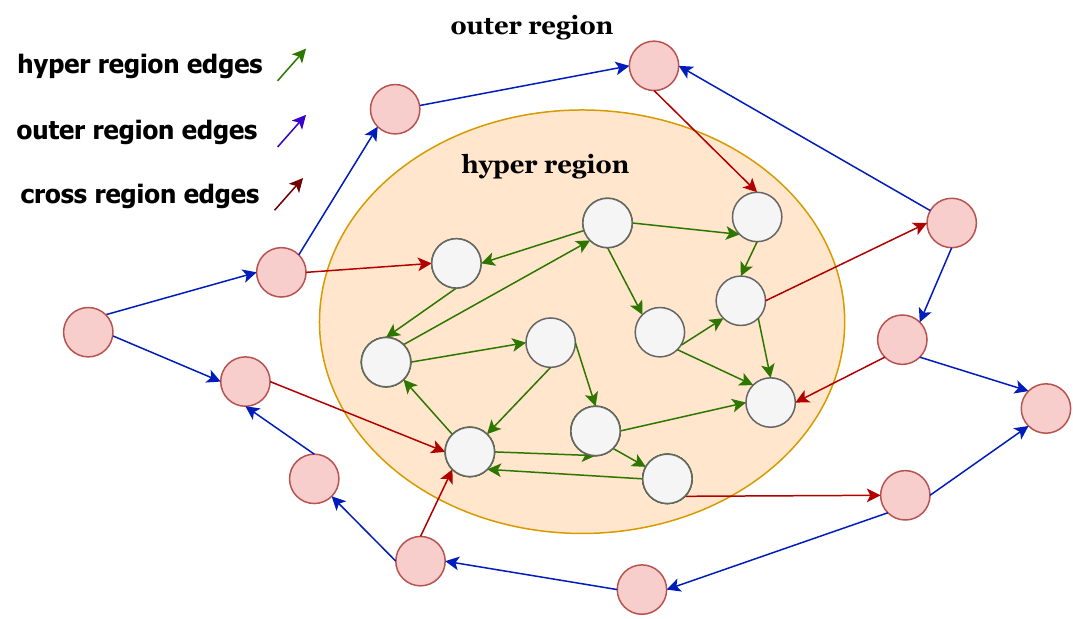}}
\caption{A hyper region in the whole DFA graph}
\label{hyper region}
\end{figure}

Hyperverse utilizes the SIMD instruction VPERMB to process the state transition, the SIMD vector of the shuffle needs to contain all the states of the DFA. While the longest SIMD vector of most modern CPUs has only 512 bits, Hyperverse can only be applied for less than 64-state DFA(the SIMD vector takes 8 bits to represent an integer). As shown in Fig.~\ref{hyper region}, to make all regexes benefit from the high performance of SIMD instruction, it is imperative to identify the hyper region within the entire DFA graph. The detection of the hyper region is far more difficult than randomly picking a region with less than 64 states. The states in the hyper region must adhere to a specific principle. When the state enters the hyper region, it tends to circulate within the region rather than promptly leaving it. This is because when a state turns to another region, it cannot benefit from the high efficiency of the SIMD instruction and requires additional operations to perform the transition. A bad hyper region results in minimal matching progress with the SIMD-based state transition algorithm and requires numerous change operations, significantly impairing the efficiency of the DFA model. 

\subsection{Good Hyper Region Properties}\label{AA}
The SIMD-based state transition algorithm has better state transition speed, so in Hyperflex, we hope once a state enters the hyper region, it is better to stay in the region as much as possible to make more use of its speed advantage, which is the principle when detecting the hyper region. To achieve that, besides containing states less than 64, a good hyper region whose states adhere to the principle should have the following three properties.

\begin{Prp1}
\label{prp:1}
States in the region better come from the same strongly connected component.
\end{Prp1}

\begin{Prp1}
\label{prp:2}
States in the region should have relatively large state stickiness.
\end{Prp1}

\begin{Prp1}
\label{prp:3}
States in the region are better located near the DFA start state.
\end{Prp1}

Next, we will introduce the three properties in detail.
\subsubsection{Property 1}
We hope the hyper region better does state transition inside itself. In graph theory, there is one concept that has similar characteristics: strongly connected component(SCC). A graph is said to be strongly connected if every vertex is reachable from every other vertex. The strongly connected components of an arbitrary directed graph form a partition into subgraphs that are themselves strongly connected. As shown in Fig.~\ref{hyper region}, the states in the same strongly connected component are more likely to transition in the inter-region, and when it leaves the region, it will never come back. Thus, states in the hyper region are better to come from the same strongly connected component as possible.
\subsubsection{Property 2}
\begin{figure}[htb]
\centerline{\includegraphics[width=0.3\textwidth]{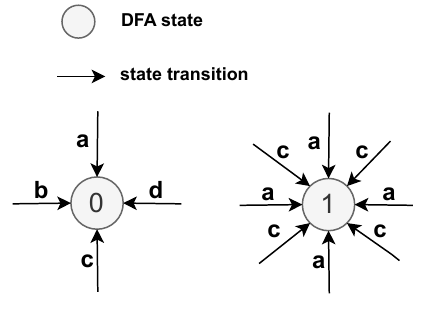}}
\caption{Two states of different in-edges to show the state stickiness}
\label{stickiness}
\end{figure}

To evaluate the different states from a strongly connected component, we propose the concept of \textit{``state stickiness''}. State stickiness represents the number of different characters on the in-edges of a state. The large stickiness means a relatively large possibility of returning to this state from adjacent states. For example, as depicted in Fig.~\ref{stickiness}, state 0 has 4 different in-edges, thus its state stickiness is 2. Meanwhile, state 1 has 8 in-edges, but its state stickiness is 2 as it only has two types \textit{`a'} and \textit{`c'}. State 0 has a relatively larger possibility of back transition to itself. Since hyper region needs a high probability of internal state transition, we think states in the region are better to have higher state stickiness.
\subsubsection{Property 3}
According to the characteristics of DFA model matching, subsequent states are more likely to revert to previous states than to progress forward. Thus, states closing to DFA starting state are more likely to be active during DFA runtime. Similarly, a hyper region closer to the DFA starting state is more likely to be used, compared to other regions far from the DFA graph entrance. Therefore, the search process should start from the strongly connected component near the DFA graph entrance.

\subsection{Hyper Region Detection}
The above three properties guide the search of the hyper region. First, through a comprehensive analysis of the DFA graph, the algorithm identifies all SCCs in the graph, and then selects an appropriate starting search region from them. This process leverages \textbf{Property 1}.

Next, the algorithm traverses each SCC in order of its distance from the DFA's initial state, summing the state stickiness values of all states within it. It then checks whether this sum exceeds the experimentally determined threshold \(\sigma\), the value of the \(\sigma\) is detailed in Part VI. If an SCC’s total state stickiness value falls below \(\sigma\), the algorithm bypasses it and continues searching for a suitable region. If no SCC meets the threshold, the algorithm terminates and defaults to the traditional state transition process. This process not only satisfies \textbf{Property 3}, which prioritizes the SCC closer to the DFA's initial state, but also leverages \textbf{Property 2}, ensuring that states within the region exhibit strong state stickiness, thereby increasing the likelihood of efficient looping within the region.

Upon identifying the starting state, the hyper region is expanded to encompass a maximum of 64 states. We choose Board Forward Search(BFS) to expand the hyper region. Because the BFS continuously involves nearby states as much as possible, the states in the hyper region are more likely to come from the same strongly connected component closer to the DFA starting state, leveraging \textbf{Property 1} and \textbf{Property 3}. After region expansion to a candidate hyper region, a scoring model is executed to decide whether to accept the hyper region. The detailed design of the evaluation model is presented below.

\subsection{Hyper Region Evaluation}
The detection of the hyper region is a qualitative analysis based on the three properties. To more accurately assess the goodness of the candidate region, we have developed a quantitative scoring model termed region leakiness.

\begin{figure}[htb]
\centerline{\includegraphics[width=0.5\textwidth]{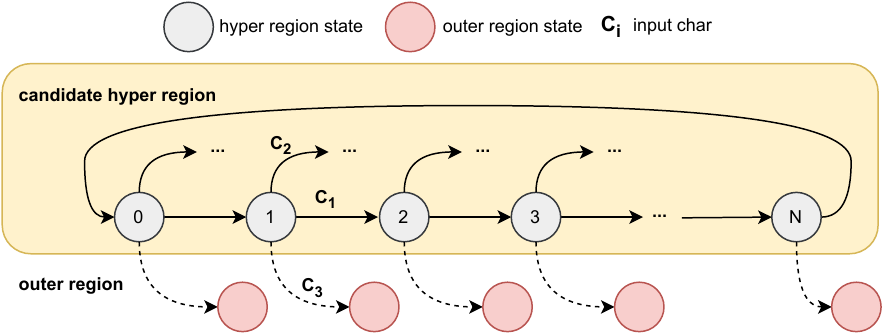}}
\caption{Leakage probabilities of a candidate hyper region}
\label{leekiness}
\end{figure}

Whenever the state transits from the hyper region into the outer region, this is referred to as a leakiness transition. In Fig.~\ref{leekiness}, the dashed edges represent leakiness transitions. To calculate the probability of a leakiness transition, we need to get the possibility of a state transition via a particular out-edge first. As Fig.~\ref{leekiness} shows, \textit{$C_1$}, \textit{$C_2$} and \textit{$C_3$} are 3 mutually exclusive character sets corresponding to 3 out-edges of state 1. $ \lvert C_1 \rvert $,$ \lvert C_2 \rvert $ and $ \lvert C_3 \rvert $ means the count of characters of each set, the value is called the width of the edge, denoted by \textit{width(e)}. In this model, the total number of characters in the ASCII set is 256, and thus the sum of all outgoing edge widths is 256. The formula of edge transition probability is as follows.  
\begin{equation}
P_{\text{transit}}(e) = \frac{\textit{width}(e)}{256}
\label{eq:1}
\end{equation}

Formula \ref{eq:1} is the probability of a state transiting to the subsequent state through a specified out-edge. This formula serves as a cornerstone for the subsequent calculation of hyper region leakiness. And the probability of the candidate hyper region leakiness is utilized to assess the quality of the region. Ideally, a high-quality hyper region should exhibit a low leakage probability. Formula \ref{eq:2} specifically addresses the leakiness probability of a given state.
\begin{equation}
P_{\text{leak}}(u) = 
\begin{cases}
\hspace{2em} 100\%, &u \in \textit{Outer Region} \\
\sum\limits_{\substack{v = \text{dest of } e \\ e \in \text{out-edge}(u)}} P_{\text{transit}}(e) \times P_{\text{leak}}(v)&u \in \textit{Hyper Region}
\end{cases}
\label{eq:2}
\end{equation}

In accordance with Formula \ref{eq:2}, the outer region state exhibits a leakage probability of 100\%. The leakage probability of a hyper region state is calculated by the weighted average of the leakiness probabilities of all its subsequent states, where the weight is the out-edge transition probability. 

The leakiness evaluation model accesses the leakiness probability of the candidate hyber region. The larger the result is, the more possible this candidate region has state transition leakage into the outer region. We treat the leakiness result of the starting state as the leakiness of the candidate region.
When the leakiness probability of a candidate hyper region is below the region leakiness threshold, $\lambda$, the region is considered a valid hyper region, and SIMD state transitions are applied. Otherwise, the DFA will use the traditional state transition algorithm. The leakiness threshold $\lambda$, is determined experimentally, and the specific experimental details will be discussed in Part VI.

\subsection{Hybrid DFA State Finalization}
Following the leakiness evaluation, a decision is made on whether to accept the candidate region as a hyper region. If not, all further matching processes are handled by the traditional DFA model. Conversely, if the region is identified as a hyper region, it becomes necessary to reconstruct the state transition table. This involves dividing the entire graph into two regions, which leads to the development of two separate DFA transition tables: table $T_t$ for outer region, and table $T_s$ for hyper region. Each transition table only contains the states from their respective region. Table $T_s$ is a one-dimensional array containing each character's transition mask vector, and table $T_t$ is a two-dimensional array like the traditional transition table, containing all the state's next transition state through each character. The building progress is as follows.

Initially, the state values within the entire DFA graph are reconfigured. A hash table $T_n$ is established to maintain the relationship between each state and its newly assigned value. The states in the hyper region are reassigned numbers ranging from $0$ to $63$ as the sequence they are added to the hyper region. Similarly, the states in the outer region are renumbered starting from $64$, corresponding to their original size sequence. In the two newly created transition tables, these reassigned numbers are utilized to represent the current state.

This design has two main advantages. Firstly, it maximizes the utilization of the SIMD mask vector space in the table $T_s$. The SIMD mask vector can only incorporate 64 states. Without reassigning state values, the mask would be unable to cover all states of the hyper region. Secondly, this strategy significantly reduces the size of the table $T_t$. Consistent with the principle of locality, reducing the array size contributes to decreasing the frequency of cache and memory swaps, leading to enhanced performance. 

% There are many grammar mistakes I am going to revise
\section{Hybrid State Transition Algorithm}
\begin{figure}[htb]
\centerline{\includegraphics[width=0.5\textwidth]{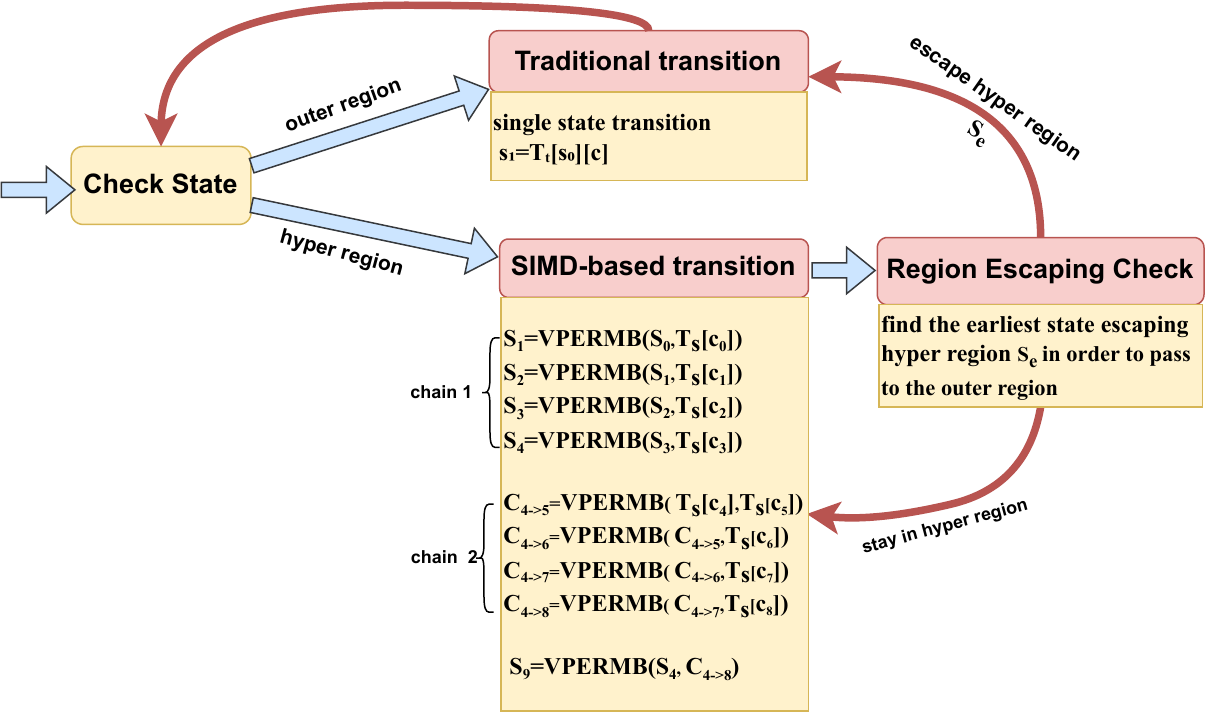}}
\caption{The process of the hybrid transition algorithm}
\label{fig_tranisition}
\end{figure}

After detecting the appropriate hyper region and completing hybrid DFA state finalization, the hybrid transition algorithm is responsible for doing each state transition process. As the Fig.~\ref{fig_tranisition}, this algorithm uses the 2 state transition tables: table $T_t$ for outer region and table $T_s$ for hyper region. It keeps checking the current state type to decide which table to use for subsequent transition.
\subsection{The Procedure of the Hybrid Transition}
When the hyper region is detected and the state transition table is reconstructed, the hybrid state transition algorithm takes responsibility for performing state transitions in different regions.  By checking which region the current state belongs to, the algorithm determines which state transition table and state transition method to use, thereby ensuring both correctness and efficiency in the transitions.

The specific state transition process is illustrated in Fig.~\ref{fig_tranisition}. In this figure and the subsequent description, VPERMB refers to the SHUFFLE instruction in the AVX512 instruction set, \(s_i\) represents a state integer, and \(S_i\) denotes a SIMD state vector consisting of 64 copies of \(s_i\). Likewise, \(c_i\) is an ASCII character, \(C_i\) is the state vector fetched from \(T_s\) using \(c_i\), and \(C_{i \to j}\) signifies applying VPERMB successively to transform \(C_i\) to \(C_j\).

Firstly, the procedure commences with an analysis of the current state. Each state in this framework correlates to a unique value in the hash table $T_n$, with hyper region states assigned lower IDs and outer region states assigned higher IDs. The boundary between these regions is defined by a $S_{limit}$, indicating the largest ID for a hyper region state. States exceeding $S_{limit}$ are processed using the table $T_t$ of outer region. In contrast, states within this threshold utilize the table $T_s$ of the hyper region. In the hyper region, a SIMD-based parallel state transition algorithm is employed, and the transition process is divided into two independent chains, referred to here as Chain~1 and Chain~2. As there are no interdependencies between these two chains, they can be executed in parallel, thereby improving computational efficiency. To further exploit the high-performance potential of the SIMD-based parallel transition algorithm and reduce the frequency of state region checks, a batch state transition mechanism is introduced. In this mechanism, \(l\) input characters are handled as a single batch. Details regarding the choice of \(l\) are discussed in the following section, and for illustrative purposes, a batch size of 9 is used as an example in the Fig.~\ref{fig_tranisition}.

When performing batch state transitions in the hyper region, it is necessary to ensure that none of the states in the batch fall outside of the hyper region. However, in actual operation, it is not always possible to keep all transitions strictly within the hyper region. In particular, there are two critical scenarios that may occur.

\begin{itemize}
\item \textbf{All results remain in hyper region:} if all state transition results stay within the hyper region, the algorithm proceeds by advancing ${S_9}$ as the initial state for the subsequent iteration. This scenario effectively exploits the optimal performance of the SIMD-based parrel state transition algorithm.

\item \textbf{One of the results transitions to outer region:} if any of the state jump out of hyper region, it is imperative to identify the earliest escaped result, denoted as ${S_j}$. Since all state transition results are obtained based on the hyper region table, the escaped $S_j$ means the outer region table should be used instead, causing subsequent results to become unreliable. We need to pass $S_j$ instead of $S_9$ to the outer region transition.
\end{itemize}

\subsection{Design of the New State Transition Table}
\subsubsection{Problem Analysis}\label{AA}
Selecting the correct table and method based on a state's region is crucial in hybrid transition algorithm. Each table contains only its regional states, thus using a mismatched table inevitably leads to errors. 

The SIMD-based parallel state transition algorithm performs 9 matches in a loop, each using the hyper region mask table. This is necessary to perform the VPERMB operation and achieve parallelism, but it assumes that no state will exceed the hyper region during the process. In reality, a state may exceed the hyper region midway through the matching processes, resulting in subsequent matching errors. This type of error is referred to as an ``out-region error".

Moreover, as shown in Fig.~\ref{fig_tranisition}, the parallel state transition is divided into two separate chains: Chain~1 and Chain~2. Chain~1 generates the state vectors \(S_1\), \(S_2\), \(S_3\), and \(S_4\). By contrast, Chain~2 computes the possible transitions for each state in response to specified input characters but discards intermediate vectors \(S_5\), \(S_6\), \(S_7\), and \(S_8\). Consequently, any out-region error in these intermediate steps cannot be detected, since the algorithm does not preserve these vectors for inspection.

\begin{figure}[htb]
\centerline{\includegraphics[width=0.5\textwidth]{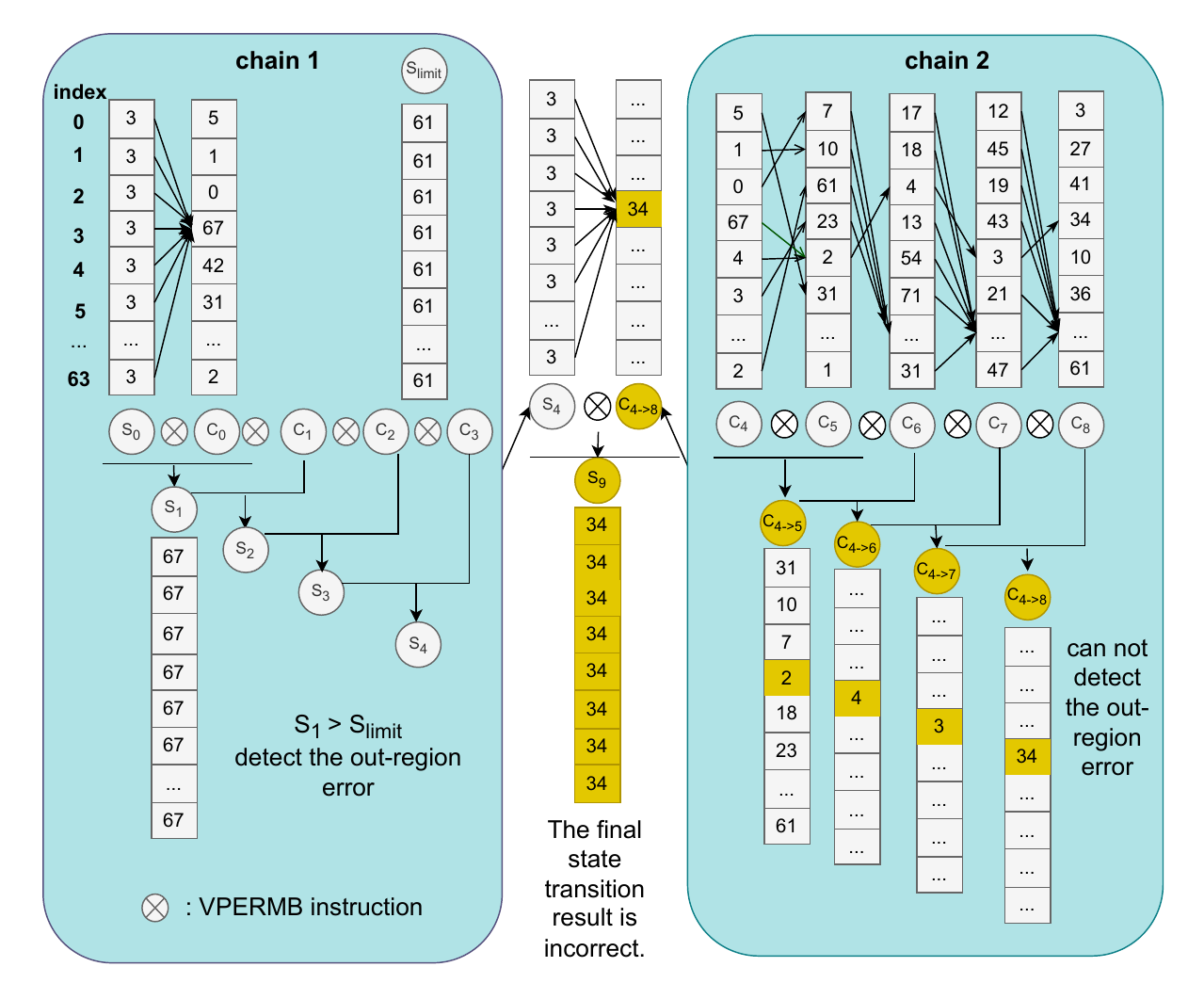}}
\caption{Out-Region Error in Chain 1 and Chain 2}
\label{gutter1}
\end{figure}

Fig.~\ref{gutter1} illustrates how the existing algorithm handles the out-region error in Chain~1 and Chain~2 respectively. If an out-region error arises in Chain~1, it is immediately detected and resolved by switching to the traditional transition algorithm. However, if the same situation occurs in Chain~2, the VPERMB instruction simply applies a modulus-based hash to the out-of-bounds index, thus masking the error. Because the resulting state vector \(S_9\) may remain below the boundary limit, it slips through subsequent boundary checks and ultimately leads to an erroneous match result.

\subsubsection{Proposed Design}
In order to address the problem above, it is imperative to detect the out-region error when it occurs. The primary challenge lies in identifying the error with the minimum possible number of comparisons. Remarkably, our design accomplishes this objective by examining only one state, thereby enhancing the efficiency of the DFA model.

To achieve this goal, we propose a new state transition table, termed the \textbf{gutter state transition table} \(T_g\), to be utilized in conjunction with the existing transition table \(T_s\). The primary distinction between \(T_g\) and \(T_s\) lies in the addition of a specific ``out-of-bounds state''. This state indicates that a certain state has exceeded the predefined boundary of the hyper region. To ensure that \(T_s\) always has sufficient capacity to store this out-of-bounds state, one regular state in the hyper region is sacrificed, reducing the maximum number of hyper region states from 64 to 63. Consequently, the value 63 is reserved as the out-of-bounds state.

\begin{figure}[htb]
  \centerline{\includegraphics[width=0.5\textwidth]{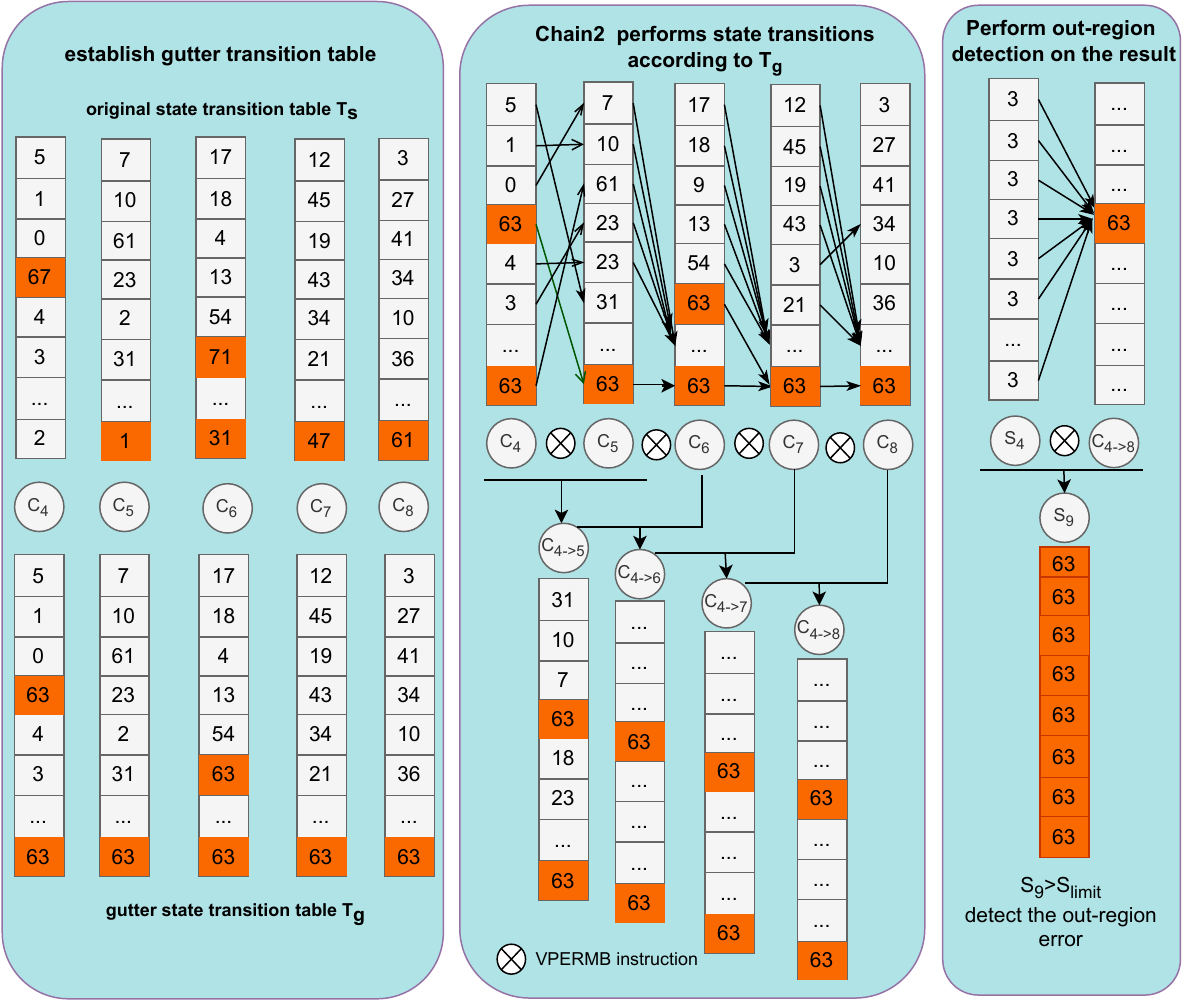}}
  \caption{Design and usage of the gutter state transition table}
  \label{fig_gutter}
\end{figure}

Fig.~\ref{fig_gutter} illustrates a concrete state transition process and highlights the use of \(T_g\). Orange squares represent state values in \(T_g\) that differ from those in the original table \(T_s\). When an out-region error occurs, the state transitions to 63 and remains there. The resulting state vector, composed entirely of 63, will always exceed \(S_{\text{limit}}\), allowing the algorithm to accurately detect the out-region error and execute appropriate error-handling or adjustments.

\subsection{Earliest Escaping State Identification Algorithm}
When an out-region error arises in Chain~1, it is necessary to identify the earliest state escaping the hyper region, enabling the outer region to manage subsequent transitions. A straightforward method compares each element of the state vectors from Chain~1 with the boundary vector \(S_{\text{limit}}\) one by one, which is intuitive but relies on conditional checks, underutilizing hardware parallelism.

To address this, we proposes a SIMD-based algorithm that identifies the earliest escaping state by grouping state vectors and performing XOR operations in parallel. This not only enhances detection speed but also supports scalable extensions to handle additional state vectors as needed.

\begin{figure}[htp]
\centerline{\includegraphics[width=0.5\textwidth]{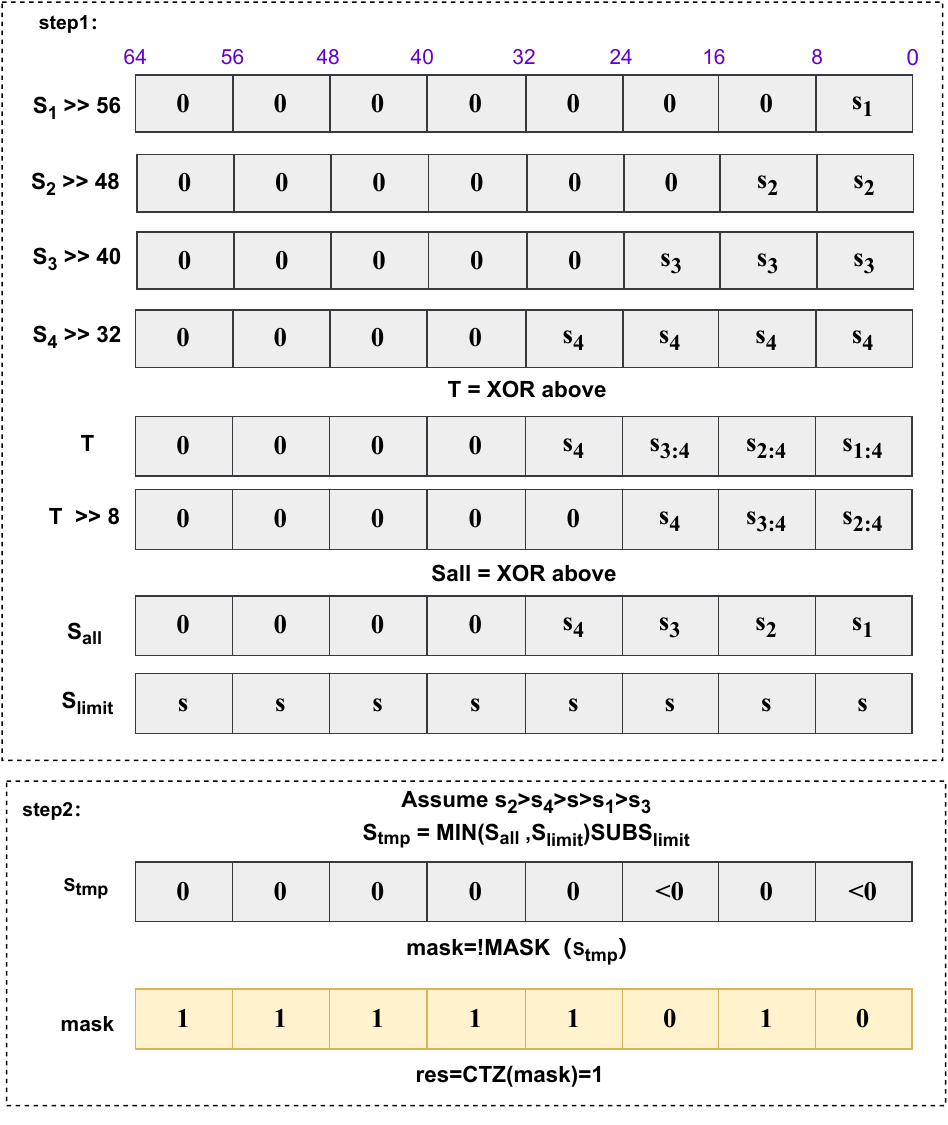}}
\caption{Workflow of the SIMD-based Escaping State Identification Algorithm}
\label{fig_two}
\end{figure}

Fig.~\ref{fig_two} illustrates the operational steps of the proposed algorithm. Although the SIMD state vector has a length of 512 bits, the algorithm processes it in 64-bit groups for simplicity, with identical operations applied to each group. Hence, only the first 64 bits are shown. Each 64-bit segment \(S_i\) consists of eight-bit state integers \(s_i\), with \(s\) denoting the maximum state value in the hyper region. The notation \(s_{1:n}\) represents successive XOR operations on \(n\) values.

First, it extracts and aggregates all state values from the Chain~1-generated vectors into a single vector. To achieve this, each state vector is shifted by a distinct offset, and the results are XORed to form an intermediate vector \textit{T}. A subsequent eight-bit right shift of \textit{T}, followed by another XOR with \textit{T}, produces the vector \(S_{\text{all}}\). In this manner, the four states from Chain~1 are consolidated and aligned in their transition order. 

Next, SIMD instructions pinpoint the earliest escaping state. Suppose \(S_2\) is the first to exceed the boundary, with \(s_2 > s_4 > s > s_1 > s_3\). The algorithm compares entries in \(S_{\text{all}}\) to \(S_{\text{limit}}\), takes their minimums, and subtracts them from \(S_{\text{limit}}\). Only states above the boundary value produce a zero in this subtraction; states below the boundary yield negative results. By extracting and inverting the sign bits, the algorithm derives an eight-bit mask that flags escaping states with 1. Finally, a \texttt{CTZ} operation locates the least significant set bit, indicating the earliest escaping state. In this example, index~1 for \(S_2\). This outcome matches expectations, demonstrating the algorithm’s ability to detect the earliest escaping state correctly.

\begin{table}[htp]
  \centering
  \caption{SIMD Instructions Utilized in the Earliest Escaping Detection Algorithm}
  \setlength{\tabcolsep}{6pt} % 调整列间距
  \renewcommand{\arraystretch}{0.9} % 调整行间距
    \begin{tabular}{c|c|c}
    \toprule
    \textbf{Symbol} & \textbf{Operation} & \textbf{SIMD Instruction} \\
    \midrule
    XOR   & Bitwise XOR         & VPXORQ   \\
    \midrule
    MIN   & Minimum Value       & VPMINUB  \\
    \midrule
    SUB   & Subtraction         & VPSUBB   \\
    \midrule
    MASK  & Extract Sign Bits   & VPTESTMD \\
    \midrule
    CTZ   & Count Trailing Zeros & VPCTZ   \\
    \bottomrule
    \end{tabular}
  \label{tab:simd-instructions}
\end{table}

\section{Evaluation}

\subsection{Environment}
We evaluate Hyperflex on a commodity CPU. We conduct experiments on a Linux server (Ubuntu 20.04) with Intel Xeon Platinum 8360Y (2.40GHz, 2 sockets, 36 cores per socket) with 256GB DDR4 memory. The processor has the support of AVX512. The hybrid DFA state transition algorithm is implemented in C. We use GCC 9.4.0 to compile them without any optimization. 
\subsection{Dataset}
To evaluate the efficiency of Hyperflex, we conduct experiments on a variety of real-world DPI regex rules and network traffic. Regex rules are collected from Snort Emerging Threats Rules v2.9.0\cite{SnortEmerging}, which is a network security rule set from the well-known DPI application Snort\cite{SnortIDS}. The input strings are derived from  from IXIA HTTP packets, Alexa non-HTTP packets, volunteer packets collected through the deployment of routers , and randomly generated byte sequences, thereby encompassing a broad spectrum of network traffic scenarios. The specific sizes of these input strings are presented in Table \ref{tab:string}.
\begin{table}[htp]
  \centering
  \caption{Input String Data}
  \label{tab:string}
  \small % 最小标准字号（约 6pt）
  \setlength{\tabcolsep}{5pt} % 最小列间距
  \begin{tabular}{c|c}
  \toprule
  \textbf{Type} & \textbf{Size} \\
  \midrule
  IXIA HTTP Packets & 122MB \\
  \midrule
  Volunteer Packets & 29MB \\
  \midrule
  Alexa non-HTTP Packets & 4MB \\
  \midrule
  Random Byte Sequences & 988KB \\
  \bottomrule
  \end{tabular}
\end{table}

\subsection{Performance Analysis}
The operation of DFA consists of two main stages: the compilation stage and the runtime stage. Accordingly, the performance evaluation experiments are divided into two parts: compilation time comparison and runtime throughput comparison. 

We select two DFA models for comparison: Mcclellan~\cite{wang2019hyperscan}, the fastest traditional DFA model in Hyperscan, and PCRE2-DFA~\cite{PCRE2}, a widely used DFA model in DPI applications. Both Mcclellan and Hyperflex are implemented within Hyperscan for experimentation. Hyperscan facilitates parallel regular expression matching by partitioning a large DFA graph into multiple subgraphs, which are independently handled by the DFA model.

Furthermore, to assess the effectiveness of the proposed region detection algorithm, we design a random region model as a baseline for comparison with Hyperflex. In this model, the start state of the hyper region is randomly selected from states near the initial DFA state, and the region expansion process does not incorporate the test of the state stickiness and leakiness of Hyperflex. This model is referred to as \textbf{HyperRandom}.

\subsubsection{Compilation Time}
During the compilation stage, DFA processing involves only the compilation of regex rules, independent of the input string. To simulate real-world  scenarios, experiments are conducted on regex rule sets of varying sizes, ranging from 10 to 1,000 rules. The regex rules used in the experiments are sourced from the Snort rule set.

\begin{table}[htp]
  \centering
  \caption{Compilation Time of Different Models (Unit: Seconds)}
  \setlength{\tabcolsep}{4pt} % 调整列间距
  \renewcommand{\arraystretch}{0.9} % 调整行间距，使表格更紧凑
  \resizebox{0.5\textwidth}{!}{ % 控制整体缩放比例
    \begin{tabular}{c|c|c|c|c}
    \toprule
    \textbf{Rules} & \textbf{Mcclellan} & \textbf{Hyperflex} & \textbf{HyperRandom} & \textbf{PCRE2-DFA} \\
    \midrule
    10   & 0.044 & 0.074 & 0.049 & 0.000059 \\
    \midrule
    50   & 0.211 & 0.264 & 0.224 & 0.000179 \\
    \midrule
    100  & 0.430 & 0.612 & 0.570 & 0.000376 \\
    \midrule
    200  & 0.810 & 1.43  & 1.25  & 0.000783 \\
    \midrule
    400  & 1.22  & 1.90  & 1.65  & 0.00144  \\
    \midrule
    600  & 2.21  & 2.51  & 2.35  & 0.00262  \\
    \midrule
    800  & 2.43  & 3.40  & 2.75  & 0.00326  \\
    \midrule
    1000 & 2.77  & 3.73  & 3.49  & 0.00403  \\
    \bottomrule
    \end{tabular}
  }
  \label{tab:compile_time_comparison}
\end{table}

The experimental results presented in Table~\ref{tab:compile_time_comparison} indicate that the compilation time of Hyperflex is slightly higher than that of Mcclellan and the randomly selected region model, HyperRandom. However, all three models exhibit significantly longer compilation times compared to PCRE2-DFA.  This is primarily due to Hyperflex's complex hyper region detection process during compilation, including tests for state stickiness and region leakiness. In contrast, Mcclellan does not incorporate region detection, and HyperRandom employs a simpler region detection algorithm, resulting in shorter compilation time. Nevertheless, all three models perform state minimization, whereas PCRE2-DFA adopts a more straightforward compilation process without extensive state compression or optimization, thereby substantially reducing its compilation time. Consequently, it exhibits inferior performance in subsequent throughput evaluations.

In practical DPI applications, the system primarily deals with large-scale network traffic, and compilation occurs only once. Consequently, the key factor determining overall system performance is runtime throughput rather than compilation time. As demonstrated in the subsequent throughput comparison, although Hyperflex incurs a slight increase in compilation time, its runtime performance gains more than compensate for this overhead. Its throughput optimization is particularly valuable in large-scale network traffic detection scenarios.

\subsubsection{Throughput}

\begin{figure*}
\centering
\subfigure[IXIA HTTP Packets]{
    \includegraphics[width=3.2in]{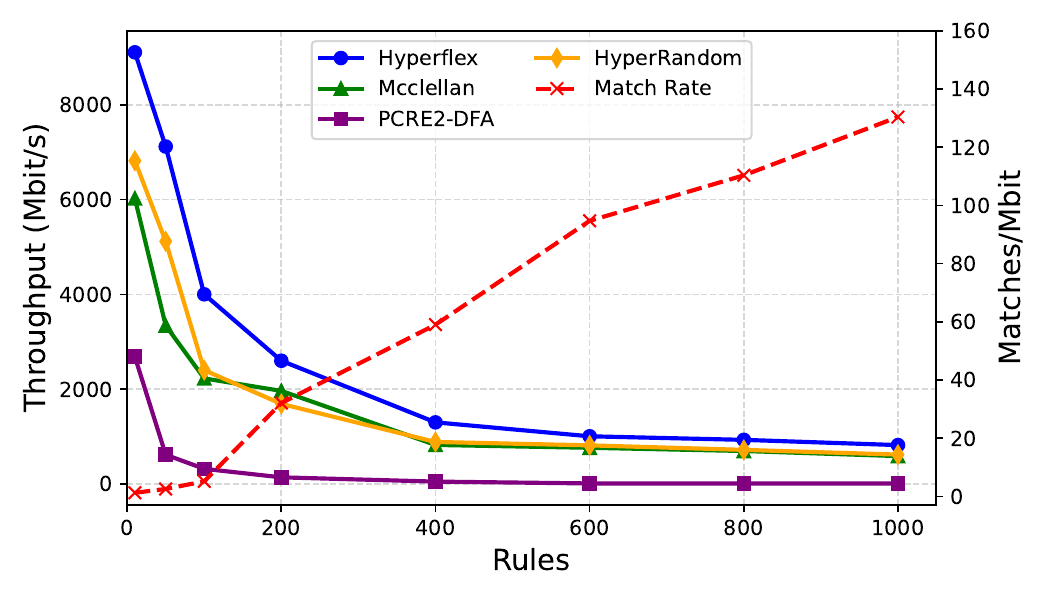}
    \label{fig_ixia}
}
\subfigure[Alexa non-HTTP Packets]{
    \includegraphics[width=3.2in]{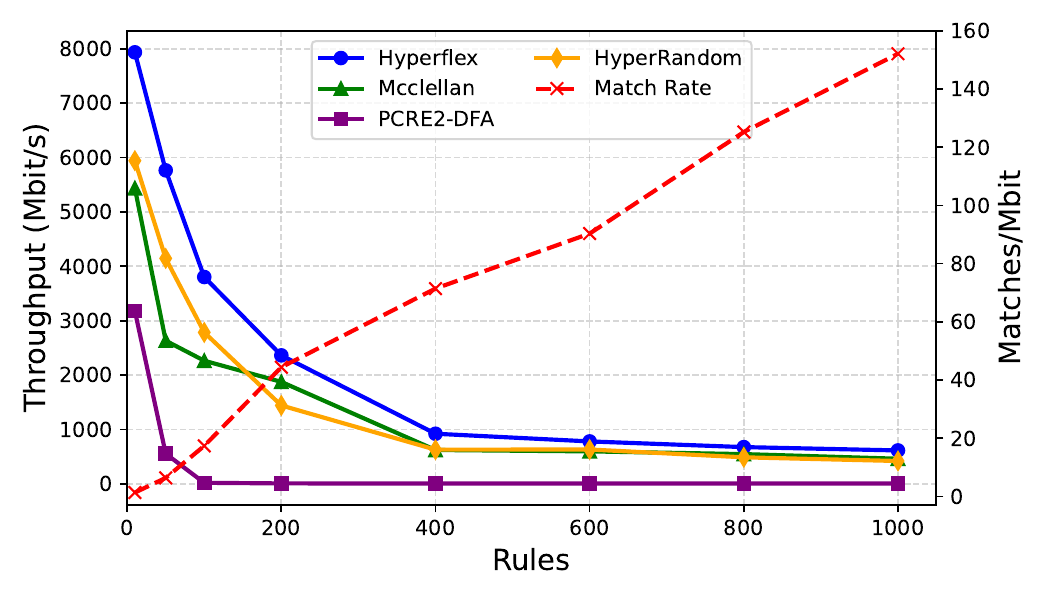}
    \label{fig_alexa}
}
\subfigure[Volunteer Packets]{
    \includegraphics[width=3.2in]{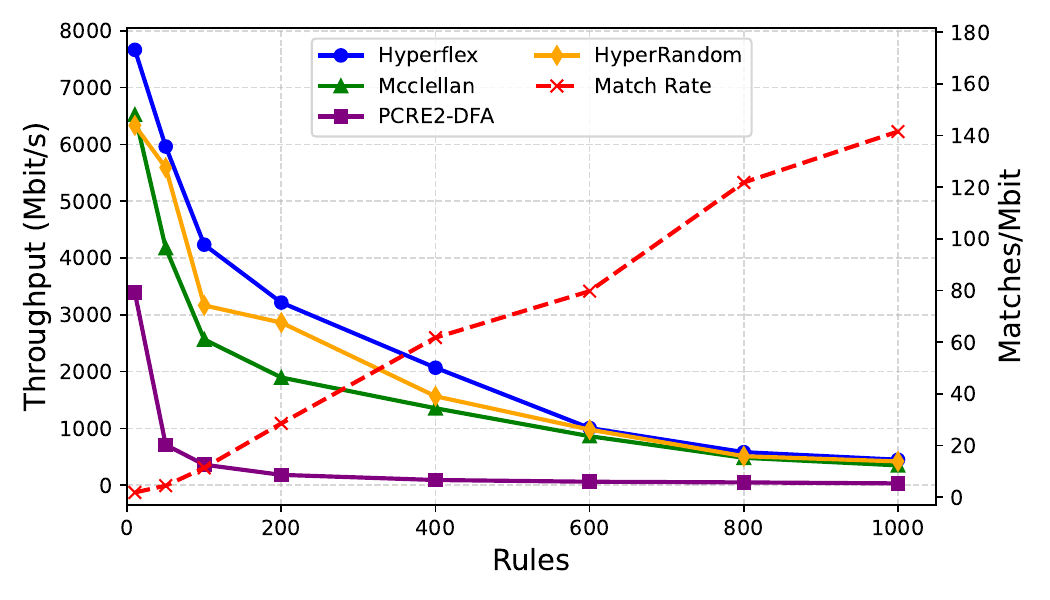}
    \label{fig_zhiyuanzhe}
}
\subfigure[Random Byte Sequences]{
    \includegraphics[width=3.2in]{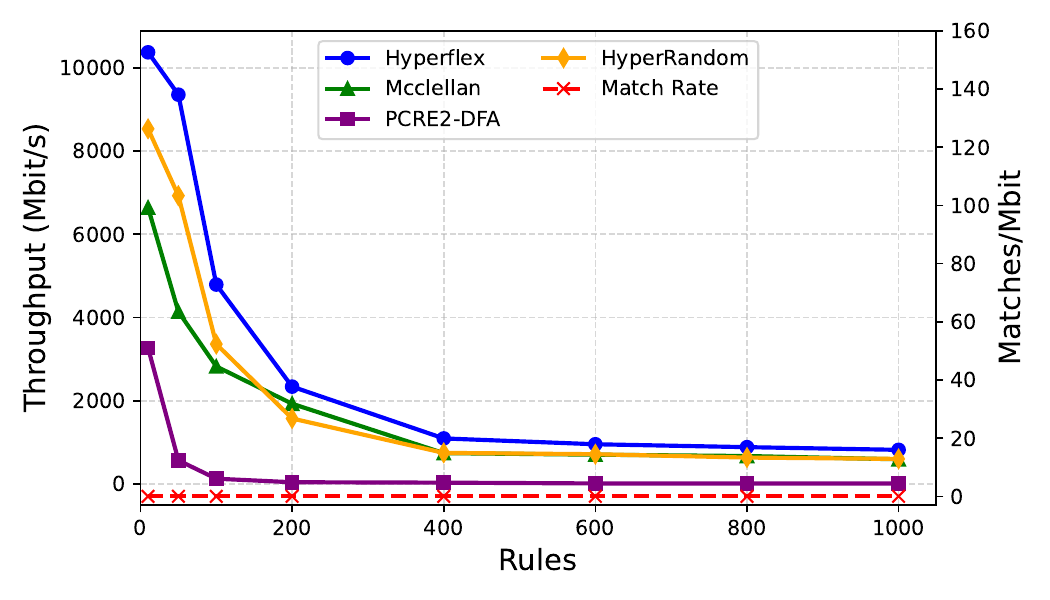}
    \label{fig_random}
}
\caption{\normalfont The throughput performance of different models among 4 types of strings}
\label{fig:throughput_comparison}
\end{figure*}

To comprehensively evaluate the runtime performance of Hyperflex, four sets of experiments are conducted using four different types of input strings, as detailed in Table~\ref{tab:string}. These experiments are performed alongside regex rules sourced from the Snort rule set, with the number of rules ranging from 10 to 1,000. The experimental results, presented in Fig.s~\ref{fig_ixia}, \ref{fig_alexa}, \ref{fig_zhiyuanzhe}, and \ref{fig_random}, consistently demonstrate that Hyperflex achieves higher throughput compared to Mcclellan, PCRE2-DFA, and HyperRandom in almost all cases.

Additionally, the match rate remains consistent across all models. For the random byte sequences input, the match rate is significantly lower. For the other three types of input strings, the match rate increases as the number of rules grows, aligning with expectations and further validating the accuracy of Hyperflex in regex matching.

In Fig.~\ref{fig_ixia}, Hyperflex achieves 1.32x $\sim$ 2.13x performance of Mcclellan, 1.23x $\sim$ 1.66x performance of HyperRandom, and 3.39x $\sim$ 91.25x performance of PCRE2-DFA. In Fig.~\ref{fig_alexa}, Hyperflex achieves 1.24x $\sim$ 1.89x performance of Mcclellan, 1.34x $\sim$ 1.64x performance of HyperRandom, and 2.49x $\sim$ 111.81x performance of PCRE2-DFA. In Fig.~\ref{fig_zhiyuanzhe}, Hyperflex achieves 1.17x $\sim$ 1.69x performance of Mcclellan, 1.07x $\sim$ 1.33x performance of HyperRandom, and 2.26x $\sim$ 71.54x performance of PCRE2-DFA. In Fig.~\ref{fig_random},  Hyperflex achieves 1.21x $\sim$ 2.27x performance of Mcclellan, 1.21x $\sim$ 1.49x performance of HyperRandom, and 3.17x $\sim$ 86.31x performance of PCRE2-DFA.

Hyperflex, Mcclellan, and HyperRandom all support parallel regex matching, while PCRE2-DFA only supports serial rule matching. Therefore, as the number of rules increases, the performance of PCRE2-DFA significantly lags behind the other models. The experimental results demonstrate that Hyperflex exhibits a clear performance advantage in regular expression matching, making it suitable for high-throughput applications such as DPI.

\subsection{Parameter Selection}
The selection of key parameters, including the state stickiness threshold $\sigma$, the region leakiness threshold $\lambda$, and the number of characters per batch $l$ for parallel state transitions, is based on experimentation. These parameters play a crucial role in determining the throughput of Hyperflex. The parameter tuning experiments use IXIA HTTP Packets as the input string and three different sets of regex rule sets of varying sizes.

\subsubsection{State Stickiness Threshold $\sigma$}
\begin{figure}[htb]
\centerline{\includegraphics[width=0.5\textwidth]{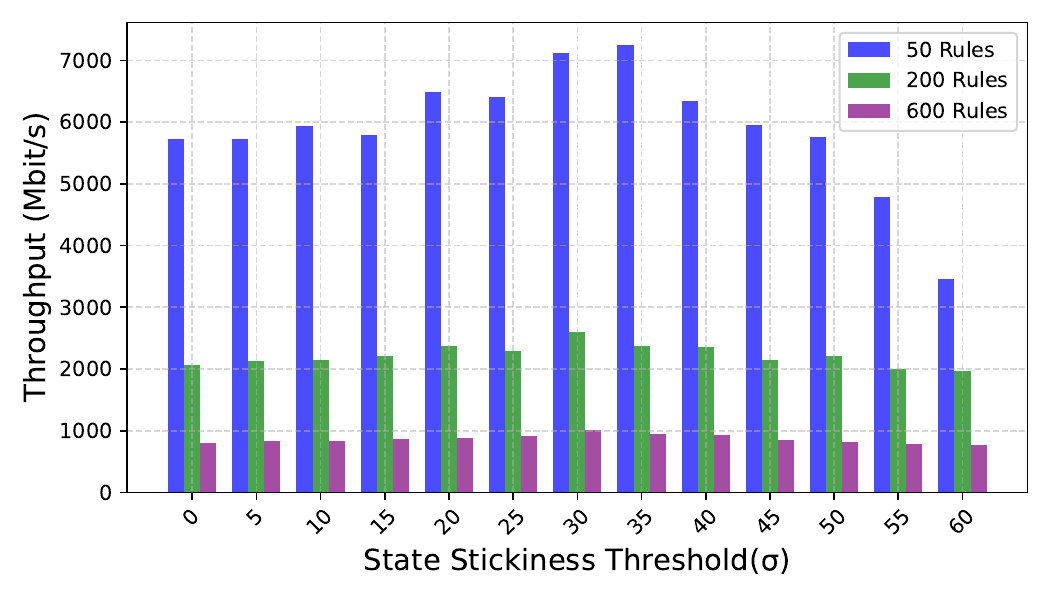}}
\caption{Throughput comparison of Hyperflex under different 
$\sigma$ values} \label{fig_huigui}
\end{figure}
State stickiness measures the propensity of states within a region to loop back to themselves during transitions. In the hyper region detection process, a strongly connected component is accepted only if its stickiness exceeds a predefined threshold; otherwise, the system checks other components. If no component meets this threshold, no hyper region is formed. Although higher thresholds make establishing hyper regions more difficult, once established, they promote frequent internal loops that better leverage SIMD-based parallel state transitions.

As shown in Fig.~\ref{fig_huigui}, when \(\sigma < 20\), the accepted strongly connected components exhibit low quality, resulting in suboptimal throughput. Conversely, when \(\sigma > 35\), the threshold is excessively high, preventing some regex rules from forming hyper region and degrading performance. Throughput is optimal when \(\sigma\) ranges from 20 to 35. Specifically, while \(\sigma = 30\) performs slightly worse than \(\sigma = 35\) under 50 regex rules, it achieves better results with larger sets of 200 and 600 rules. Given that practical DPI scenarios often involve large-scale rule sets, we set the threshold to 30 by default.

\subsubsection{Leakage Probability Threshold $\lambda$}
\begin{figure}[htb]
\centerline{\includegraphics[width=0.5\textwidth]{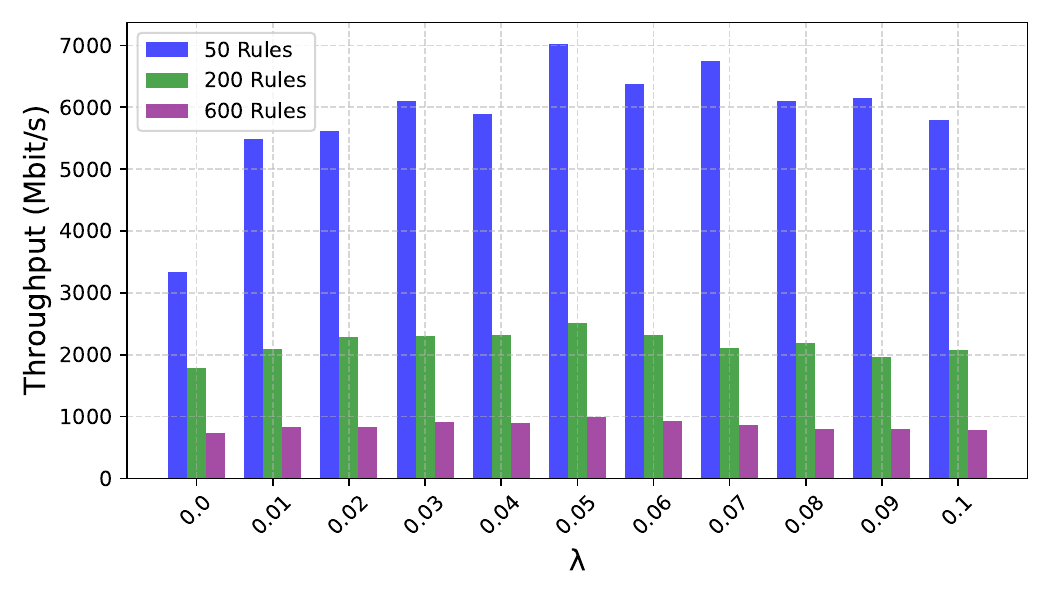}} 
\caption{Throughput comparison of Hyperflex under different $\lambda$ values}
\label{yuejie}
\end{figure}
The region leakiness threshold $\lambda$ is used to evaluate the quality of the candidate hyper region. It determines whether a candidate hyper region is accepted and processed using the SIMD-based parallel state transition algorithm or rejected in favor of the traditional DFA state transition algorithm. When the threshold is set too low, only a few regions can be accepted as accelerated regions. Conversely, if the threshold is set too high, low-quality regions may also be accepted. Therefore, it is necessary to experimentally determine an appropriate threshold.

As shown in Fig.~\ref{yuejie}, Hyperflex performs poorly when \(\lambda = 0\), because no candidate regions pass the leakiness evaluation, preventing any benefit from SIMD-based parallel state transitions. As \(\lambda\) increases, more candidate accelerated regions meet the quality criteria, thereby improving throughput. However, once \(\lambda\) exceeds 0.05, overall throughput declines due to the acceptance of low-quality regions, whose frequent exits from hyper region introduce time overhead that negatively impacts overall performance. Experimental results indicate that \(\lambda = 0.05\) optimizes Hyperflex’s performance across three different scales of regex rule sets.

\subsubsection{Number of Characters per Batch $l$ for Parallel State Transitions}

\begin{figure}[H]
\centerline{\includegraphics[width=0.5\textwidth]{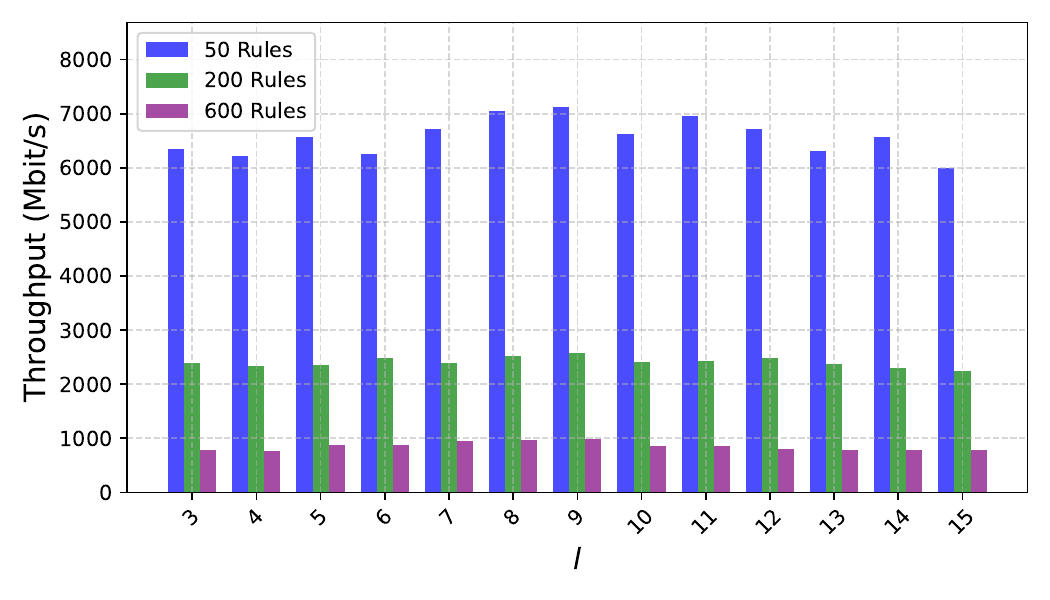}} 
\caption{Throughput comparison of Hyperflex under different $l$ values} \label{fig_depth} 
\end{figure}

The value of $l$ not only affects the number of calls to the region identification algorithm but also the number of state transitions that need to be re-executed when a state of the batch escapes hyper region.  As shown in Fig. \ref{fig_depth},when $l$ increases, the number of characters processed per batch increases, reducing the number of calls to the region identification algorithm, which is beneficial for overall performance. However, as $l$ continues to grow, the computational overhead from re-executing state transitions after the earliest state escapes hyper region increases, leading to a more significant negative impact on performance. For the three different scales of regex rule sets, Hyperflex achieves optimal performance when $l = 9$.

\subsection{Performance Comparison of the Earliest Escaping State Identification Algorithm}
Based on theoretical analysis and experimental results, the primary performance improvement of Hyperflex stems from replacing traditional state transitions with the SIMD-based parallel state transition algorithm within accelerated regions. Additionally, the careful design of SIMD instructions to replace simple program statements has also contributed to performance enhancements.

The SIMD-based earliest boundary violation state identification algorithm proposed in Part V, referred to as the``SIMD identification algorithm", demonstrates significant performance improvements over the ``traditional identification algorithm". The latter sequentially compares multiple state vectors with boundary vectors to determine the earliest escape state.

\begin{figure}[htb]
\centerline{\includegraphics[width=0.5\textwidth]{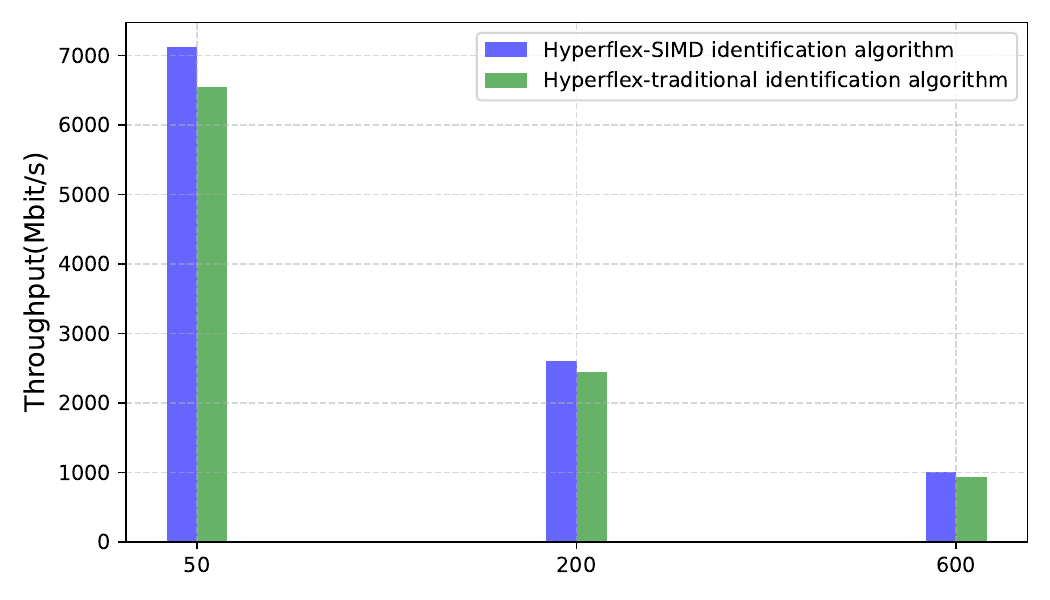}} 
\caption{Throughput comparison of Hyperflex using earliest escaping state
identification algorithm} 
\label{fig_find} 
\end{figure}

To evaluate the contribution of this optimized algorithm to the overall throughput of Hyperflex, experiments were conducted using IXIA packets as the input string, with regex rule sets of 50, 200, and 600 rules. The experimental results, as shown in Fig.~\ref{fig_find}, indicate that Hyperflex using the SIMD identification algorithm outperforms the version using the traditional identification algorithm, with an average throughput improvement of 7.5\%. This result demonstrates that the SIMD identification algorithm effectively reduces the time required to identify the earliest state that escapes from the hyper region and enhances the overall matching efficiency of the model.

\section{Conclusion}
In this paper, we propose Hyperflex, a highly-optimized,
SIMD-based DFA model, which achieves a significant
improvement in processing speed compared to the traditional
DFA model. It employs a novel region detection algorithm to
identify the region that is most suitable for
employing the SIMD-based state 
transition algorithm of the whole DFA graph. Furthermore, it introduces
an efficient hybrid transition algorithm to make different regions
under different algorithms and ensure seamless state transitions
across different regex regions. We implemented this model
on commodity CPU and conducted experiments against real-world
network traffic and DPI regex rules. Our comparative analysis
with Mcclellan, Hyperscan’s traditional DFA matching
model, demonstrates that Hyperflex achieves a throughput of
8.89Gbit/s, up to 2.27x compared with the performance of
Mcclellan. This model has been successfully integrated into
Hyperscan.

\section*{Acknowledgments}
This work was supported by National Key R\&D Program of China under Grant No. 2022YFB3102902.

% or


\ifCLASSOPTIONcaptionsoff
  \newpage
\fi



\bibliographystyle{IEEEtran}
\bibliography{IEEEabrv,main}
\begin{IEEEbiography}[{\includegraphics[width=1in,height=1.25in,clip,keepaspectratio]{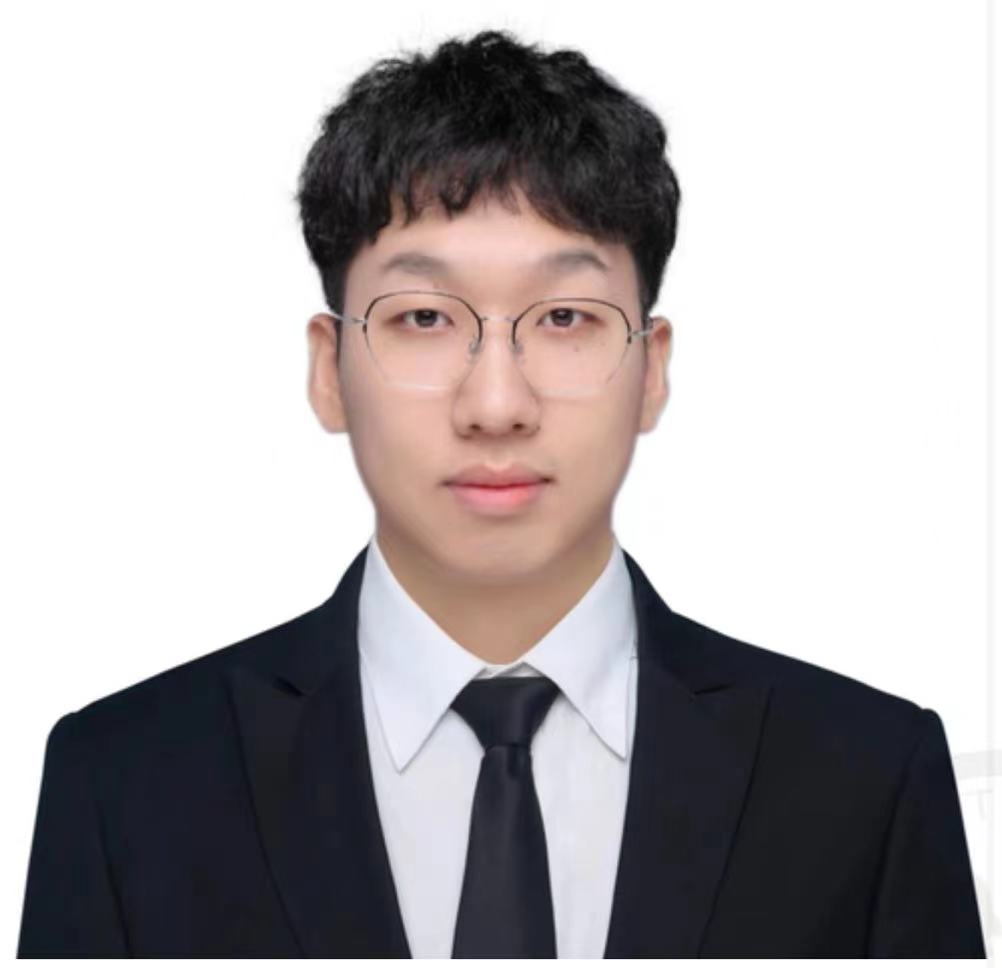}}]{Yang Liu} received his bachelor's degree in mechanical engineering from Shanghai Jiao Tong University, Shanghai, China, in 2021. He is currently working toward the master's degree with the the School of Computer Science, Fudan University, Shanghai, China. His research interests include software-defined networking and deep packet inspection.
\end{IEEEbiography}

\begin{IEEEbiography}[{\includegraphics[width=1in,height=1.25in,clip,keepaspectratio]{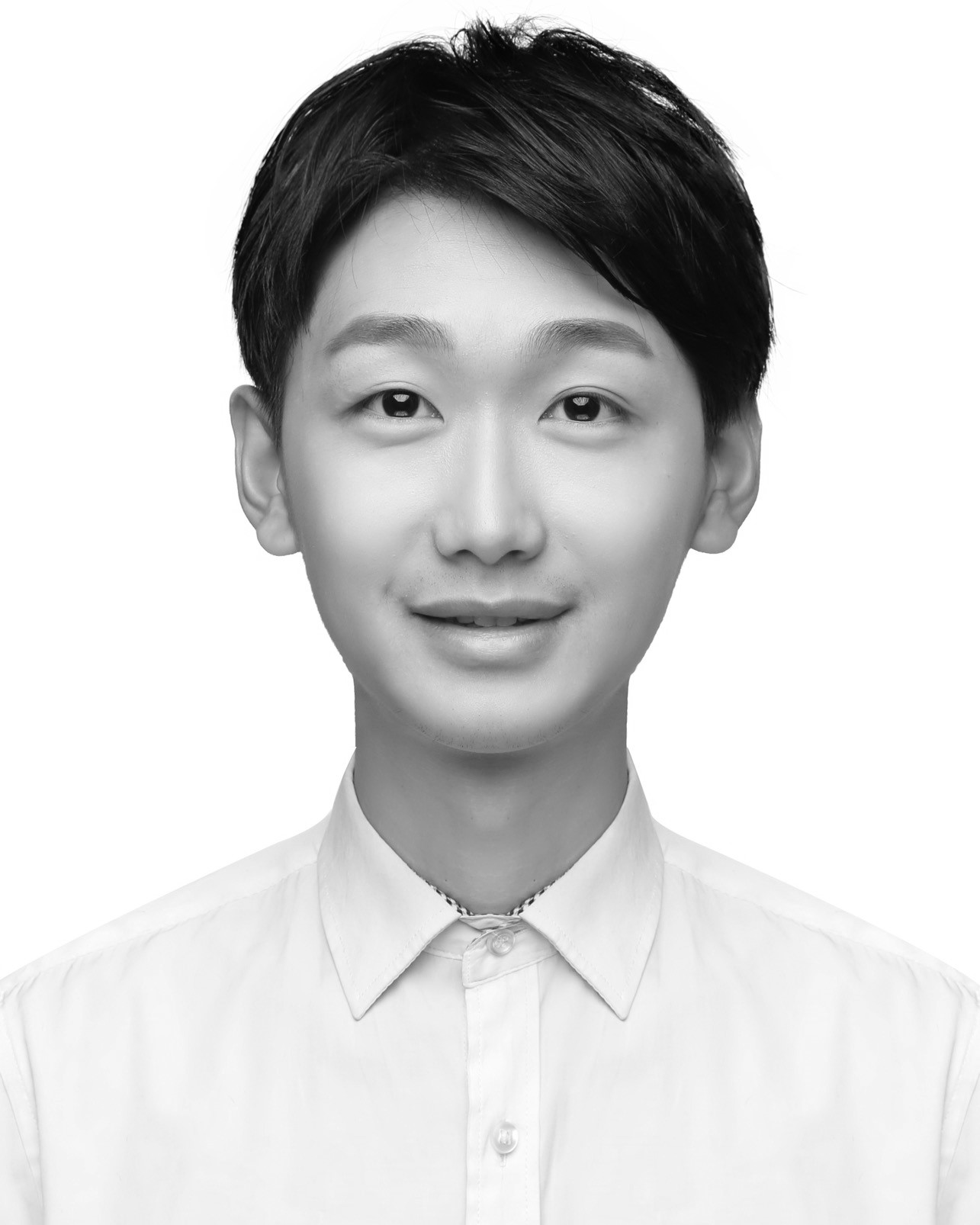}}]{Wenjun Zhu} received his master's Degree from  Nanjing University of Posts and Telecommunications in 2020. He works for Intel as a software engineer from 2020 to 2023. Now he joined Fudan University in 2023 as a software engineer. His research interests include computer networks and computer architecture.
\end{IEEEbiography}

\begin{IEEEbiography}[{\includegraphics[width=1in,height=1.25in,clip,keepaspectratio]{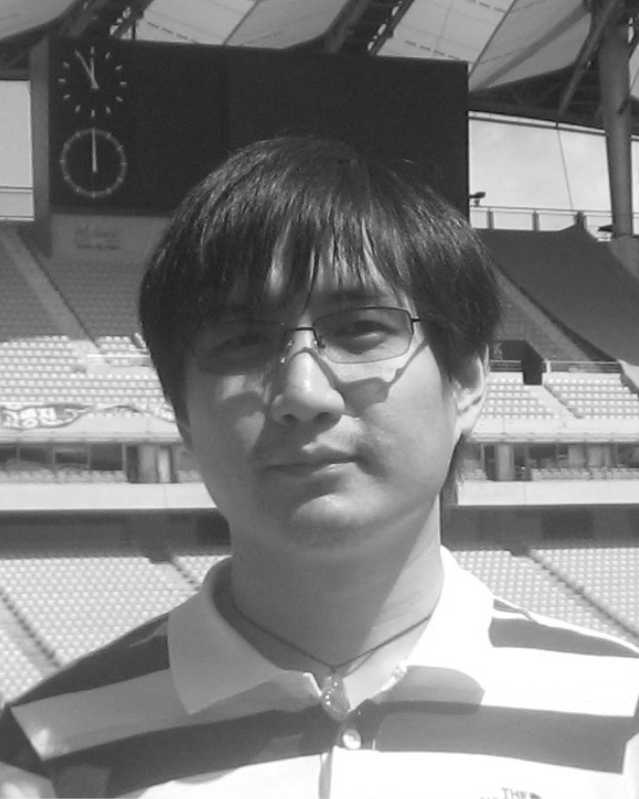}}]{Harry Chang} received his B.Eng. Degree from Shanghai Jiao Tong University in 2006. He now works for Intel as a senior software engineer. His research/work interests include computer network and performance optimization on Intel architecture.
\end{IEEEbiography}

\begin{IEEEbiography}[{\includegraphics[width=1in,height=1.25in,clip,keepaspectratio]{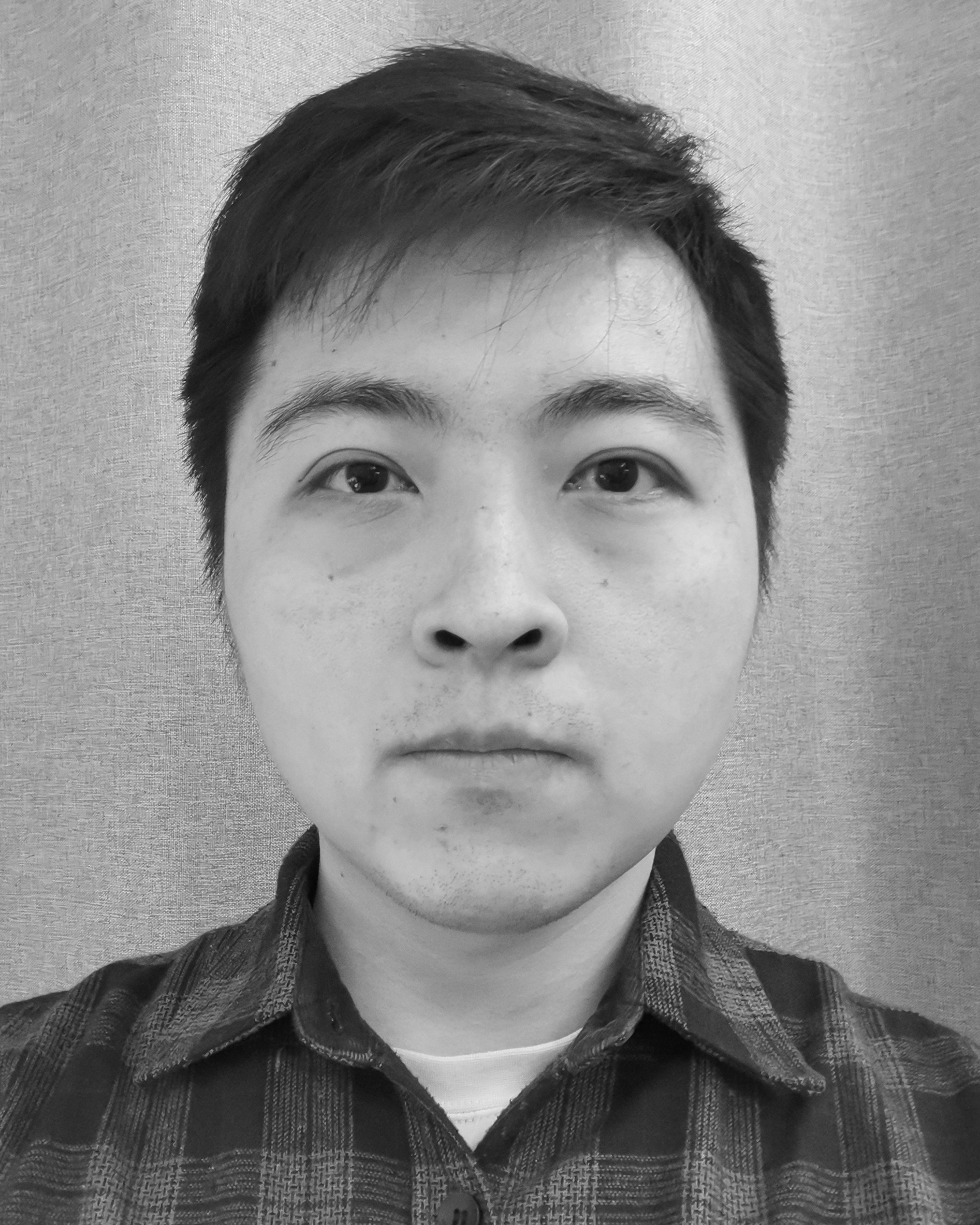}}]{Yang Hong} received his B.S. Degree from Nanjing University in 2014, and received the M.E. Degree from Shanghai Jiao Tong University in 2017. He now works for Intel as a software engineer. His work interests include SIMD algorithm optimization and pattern matching.
\end{IEEEbiography}
\vspace{4cm}
\begin{IEEEbiography}[{\includegraphics[width=1in,height=1.25in,clip,keepaspectratio]{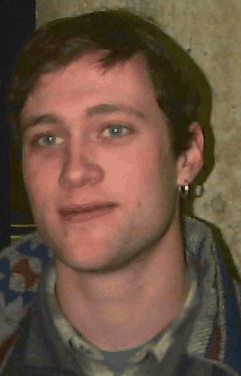}}]{Geoff Langdale} received his Ph.D. degree in Computer Science from Carnegie Mellon University. He served as the chief architect of Hyperscan, a high-performance regular expression matching engine. He is interested in writing code at the lowest architectural level and trying to pack every cycle with useful work.
\end{IEEEbiography}

\begin{IEEEbiography}[{\includegraphics[width=1in,height=1.25in,clip,keepaspectratio]{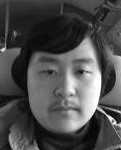}}]{Kun Qiu} received his B.Sc. Degree from Fudan University in 2013, and received his Ph.D. Degree from Fudan University in 2019. He works for Intel as a software engineer from 2019 to 2023. Now he joined Fudan University in 2023 as an assistant professor. His research interests include computer networks and computer architecture. He is a senior member of IEEE, CCF and member of ACM.
\end{IEEEbiography}

\begin{IEEEbiography}[{\includegraphics[width=1in,height=1.25in,clip,keepaspectratio]{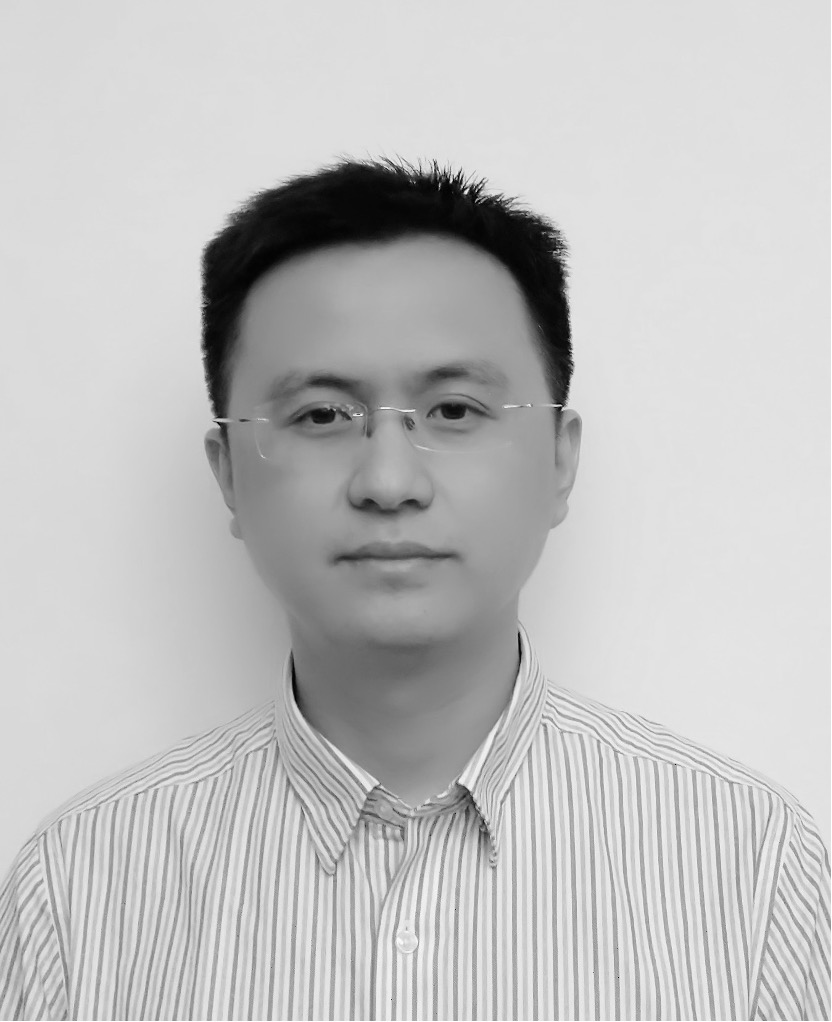}}]{Jin Zhao} received the B.Eng. degree in computer communications from Nanjing University of Posts and Telecommunications, China, in 2001, and the Ph.D. degree in computer science from Nanjing University, China, in 2006. He joined Fudan University in 2006. His research interests include software defined networking and distributed machine learning. He is a senior member of IEEE.
\end{IEEEbiography}


%\end{IEEEbiography}


%\newpage



%\vfill


\end{document}